\input ./style/arxiv-general.cfg
\documentclass[MSNbibl,number,citesort,seceqn,dvips]{arxbj}
\makeatletter
   \@ifpackageloaded{graphicx}{}{\usepackage{graphicx}}
\makeatother

\volume{23}
\issue{1}
\pubyear{2017}
\firstpage{432}
\lastpage{458}
\doi{10.3150/15-BEJ748}
\docsubty{FLA}

\makeatletter

\newtheorem{theorem}{Theorem}[section]

\newtheorem{corollary}{Corollary}[section]

\newproclaim{definition}{Definition}[section]
\newremark{example}{Example}[section]

\newtheorem{lemma}{Lemma}[section]

\newtheorem{proposition}{Proposition}[section]
\newremark{remark}{Remark}[section]

\newproclaim{assumption}{Assumption}[section]
\makeatother

\begin{document}
\begin{frontmatter}

\title{On magnitude, asymptotics and duration of drawdowns for L\'evy models}
\runtitle{On magnitude, asymptotics and duration of drawdowns for L\'
{e}vy models}

\begin{aug}
\author[A]{\inits{D.}\fnms{David}~\snm{Landriault}\thanksref{A,e1}\ead[label=e1,mark]{dlandria@uwaterloo.ca}},
\author[A]{\inits{B.}\fnms{Bin}~\snm{Li}\corref{}\thanksref{A,e2}\ead[label=e2,mark]{bin.li@uwaterloo.ca}}
\and
\author[C]{\inits{H.}\fnms{Hongzhong}~\snm{Zhang}\thanksref{C}\ead
[label=e3]{hzhang@stat.columbia.edu}}
\address[A]{Department of Statistics and Actuarial Science,
University of Waterloo, Waterloo, ON, N2L 3G1, Canada.\\ \printead{e1,e2}}
\address[C]{Department of Statistics, Columbia University, New
York, NY, 10027, USA.\\ \printead{e3}}
\end{aug}

%
\received{\smonth{6} \syear{2014}}
%
\revised{\smonth{3} \syear{2015}}

%
\begin{abstract}
This paper considers magnitude, asymptotics and duration of drawdowns
for some
L\'{e}vy processes. First, we revisit some existing results on the magnitude
of drawdowns for spectrally negative L\'{e}vy processes using an approximation
approach. For any spectrally negative L\'{e}vy process whose scale functions
are well-behaved at $0+$, we then study the asymptotics of drawdown quantities
when the threshold of drawdown magnitude approaches zero. We also show that
such asymptotics is robust to perturbations of additional positive compound
Poisson jumps. Finally, thanks to the asymptotic results and some
recent works
on the running maximum of L\'{e}vy processes, we derive the law of
duration of
drawdowns for a large class of L\'{e}vy processes (with a general spectrally
negative part plus a positive compound Poisson structure). The duration of
drawdowns is also known as the ``Time to
Recover'' (TTR) the historical maximum, which is a widely
used performance measure in the fund management industry. We find that
the law
of duration of drawdowns qualitatively depends on the path type of the
spectrally negative component of the underlying L\'{e}vy process.
\end{abstract}

%
\end{frontmatter}

\section{Introduction}\label{sec1}

Drawdowns relate to an investor's sustained loss from a market peak. It
is one
of the most frequently quoted indices for downside risks in the fund
management industry. Drawdown quantities appear in performance measures such
as the Calmar ratio, the Sterling ratio, the Burke ratio, and others; see,
for example, Schuhmacher and Eling \cite{SE11} for a collection of such
drawdown-based performance measures. Furthermore, drawdown problems
have drawn
considerable theoretical and practical interest in various research areas
including probability, finance, risk management, and statistics; see
Section~\ref{sec1.1} for a brief literature review.

In this paper, we consider a one-dimensional L\'{e}vy process $X=\{X_{t}
,t\geq0\}$ defined on $(\Omega,\mathcal{F},\mathbf{F}=\{
\mathcal{F}%
_{t},t\geq0\},\mathbb{P})$, a filtered probability space satisfying
the usual
conditions. The drawdown process of $X$ is defined as
\[
Y_{t}=M_{t}-X_{t},\qquad t\geq0,
\]
where $M_{t}=\sup_{0\leq u\leq t}X_{u}$ is the running maximum (historical
peak) of $X$ at time $t$. Let
\[
\tau_{a}=\inf \{ t\geq0:Y_{t}>a \},
\]
be the first time the magnitude of drawdowns exceeds a pre-specified threshold
$a>0$. Given that $ ( \max_{0\leq s\leq t}Y_{s}>a ) = (
\tau_{a}<t ) $ $\mathbb{P}$-a.s., the distributional study of the
maximum drawdown in magnitude is equivalent to the study of the
stopping time
$\tau_{a}$.

However, from a risk management standpoint, the magnitude itself is not
sufficient to provide a comprehensive risk evaluation of extreme drawdown
risks. For instance, for extreme risks such as tornado and flooding, it is
natural to also investigate the frequency and the duration of drawdowns.
Landriault \textit{et al.} \cite{LLZ15} recently studied the frequency of
drawdowns for
a Brownian motion process by defining two types of drawdown time sequences
depending on whether a historical running maximum {is reset or not}. In this
paper, we will consider the duration of drawdowns, also known as
``Time to Recover'' (TTR) the historic
running maximum in the fund management industry.

Mathematically, the duration of drawdowns of a stochastic process can be
considered as the length of excursions from its running maximum. For
$t\geq0$,
let $G_{t}:=\sup\{0\leq s\leq t:Y_{s}=0\}$ be the last time the
process $Y$ is
at level $0$ (or equivalently $X=M$) before or at time $t$. The drawdown
duration at time $t$ is therefore $t-G_{t}$. We then define a stopping time
%
\begin{equation}
\eta_{b}=\inf \{ t\geq b:t-G_{t}\geq b \},
\label{eta b}%
\end{equation}
that is the first time the duration of drawdowns exceeds a
pre-specified time
threshold $b>0$. Equivalently, the event $(\eta_{b}>t)$ implies that the
maximum duration of drawdowns before time $t$ is shorter than $b$.

The stopping time $\eta_{b}$ is related to the so-called Parisian
time, which
is the first time the length of excursions from a fixed spatial level (rather
than its running maximum) exceeds a pre-specified time threshold; see,
for example,
Chesney \textit{et al.} \cite{CJY97} and Czarna and Palmowski \cite{CP11}. Further,
Loeffen \textit{et al.} \cite{LCP13} provided an unified proof to derive the
probability that the Parisian time occurs in an infinite time horizon (known
as the Parisian ruin probability in actuarial science) for spectrally negative
L\'{e}vy processes. Notice that, in contrast to the Parisian time, the
stopping time $\eta_{b}$ is almost surely finite (e.g., page 105 of Bertoin
\cite{B96}), which motivates us to study the Laplace transform (LT) of
$\eta_{b}$ in this paper. Another related concept is the so-called red period
of the insurance surplus process; see Kyprianou and Palmowski \cite
{KP07}. The
red period corresponds to the length of time an insurance surplus process
shall take to recover its deficit at ruin. But it is different than the
distributional study of $\eta_{b}$, especially when $X$ has no
negative jumps
(e.g., Brownian motion).

\subsection{Literature review on drawdowns}\label{sec1.1}

Taylor \cite{T75} first derived the joint Laplace transform of $\tau
_{a}$ and
$M_{\tau_{a}}$ for Brownian motion processes. Later on, it was
generalized by
Lehoczky \cite{L77} to time-homogeneous diffusion processes. Douady \textit{et al.}
\cite{DSY00} and Magdon \textit{et al.} \cite{MAPA04} derived infinite series
expansions for the distribution of $\tau_{a}$ for a standard and drifted
Brownian motion, respectively. For spectrally negative L\'{e}vy processes,
Mijatovi\'{c} and Pistorius~\cite{MP12} obtained a general sextuple
formula for
the joint Laplace transform of $\tau_{a}$ and the last passage time at level
$M_{\tau_{a}}$ prior to $\tau_{a}$, together with the joint
distribution of
the running maximum, the running minimum, and the overshoot of $Y$ at
$\tau_{a}$. Also, an extensive body of literature exists on {the dual of
drawdowns, drawups, which measure the increase in value of an underlying
process from its running minimum; see, for instance, }Pistorius \cite{P04},
Hadjiliadis and Ve\v{c}e\v{r} \cite{HV06}, Pospisil \textit{et al.} \cite{PVH09}, and
Zhang and
Hadjiliadis \cite{ZH10,ZH12}.

In finance and risk management, researchers have devoted considerable effort
in assessing, managing, and reducing drawdown risks. For instance, Grossman
and Zhou \cite{GZ93} examined a portfolio selection problem subject to
drawdown constraints. Cvitanic and Karatzas \cite{CK95} extended the
discussion to multiple assets. Chekhlov \textit{et al.} \cite{CUZ05} proposed a new
family of risk measures and studied problems of parameter selection and
portfolio optimization under the new measures. Pospisil and Ve\v{c}e\v{r} \cite{PV10}
invented a new class of Greeks to examine the sensitivity of investment
portfolios to drawdowns. Carr \textit{et al.} \cite{CZH11} designed some European-style
digital drawdown insurance contracts and proposed semi-static hedging
strategies using barrier options and vanilla options. Other recent
works on
drawdown insurance are Zhang \textit{et al.} \cite{Z15,ZLH13}, among others.

In addition, a few a priori unrelated problems in finance and insurance are
also closely connected to the drawdown problematic. For instance, the pricing
of Russian options (e.g., Shepp and Shiryaev \cite{SS93}, Asmussen \textit{et al.}
\cite{AAP04} and Avram \textit{et al.} \cite{AKP04}), and the optimal dividend models
with ``reflecting barriers'' (e.g.,
Avram et
al. \cite{APP07}, Kyprianou and Palmowski \cite{KP07}, and Loeffen~\cite{L08})
are two common examples.

\subsection{Objective and structure}

In this article, we begin by developing an approximation technique in the
spirit of Lehoczky \cite{L77} to revisit several known LT results on the
magnitude of drawdowns of spectrally negative L\'{e}vy processes via basic
fluctuation identities.

Second, as the threshold of drawdown magnitude $a\downarrow0$, we
examine the
asymptotic behavior of those LTs for any spectrally negative L\'{e}vy process
whose scale functions are well-behaved at $0+$ (see Assumption~\ref{assume1}
below). We also show that such asymptotics are robust with respect to the
perturbation of arbitrary positive compound Poisson jumps, and hence obtain
the asymptotics of drawdown estimates for a class of L\'{e}vy models with
two-sided jumps.

Finally, we study the duration of drawdowns via the LT of $\eta_{b}$. First,
an approximate scheme for the LT of $\eta_{b}$ is developed. To obtain a
well-defined limit, we turn our problem to the behavior of the density
of the
running maximum process $M$ and the convergence of some potential
measure of
the drawdown process $Y$. Thanks to the asymptotic results obtained and some
recent works on the distribution of the running maxima of L\'{e}vy processes
(e.g., Chaumont \cite{C13}, Chaumont and Ma{\l}ecki \cite{CM13}, and
Kwa\'{s}nicki \textit{et al.} \cite{KMR13}), we obtain the law of $\eta_{b}$
in terms
of the right tail of the ascending ladder time process for a class of
L\'{e}vy
process with two-sided jumps (a general spectrally negative part plus a
positive compound Poisson structure).

The rest of the paper is organized as follows. In Section~\ref
{sec:pre}, we
review the scale function of spectrally negative L\'{e}vy processes and the
ascending ladder process of a general L\'{e}vy process. In Section~\ref
{sec:mag}, we revisit some known LT results on the magnitude of drawdowns
of spectrally negative L\'{e}vy processes based on an approximation approach.
The asymptotic behavior of these LTs for small threshold is studied in
Section~\ref{sec:asym}, where we also examine the asymptotic behavior
in the presence
of positive compound Poisson jumps. In Section~\ref{sec:ddd}, the LT of
$\eta_{b}$ is derived for a large class of L\'{e}vy processes with two-sided
jumps. Some explicit examples are presented in Section~\ref{sec:ex}. For
completeness, some results on the extended continuity theorem are
presented in the \hyperref[sec:app]{Appendix}.

\section{Preliminaries}
\label{sec:pre}

In this section, we briefly introduce some preliminary results for L\'{e}vy
processes. Readers are referred to Bertoin \cite{B96} and Kyprianou
\cite{K06}
for a more detailed background.

For ease of notation, throughout the paper, we let $\mathbb
{R}=(-\infty
,\infty)$, $\mathbb{R}_{+}=[0,\infty)$ and $\mathbb{H}^{+}=\{s\in
\mathbb{C}:\operatorname{Re}(s)\geq0\}$. We denote by $\mathbb
{P}_{x}$ the law
of a L\'{e}vy process with $X_{0}=x\in%
\mathbb{R}
$. For brevity, we write $\mathbb{P}=\mathbb{P}_{0}$. The minimum of real
numbers $u,v$ is denoted by $u\wedge v=\min\{u,v\}$. For a function
$f(\cdot)$
on $(0,\infty)$ and $x_{0}\in{}[0,\infty]$, we write
$f(x)=o(g(x))$ as
$x\rightarrow x_{0}$ for a positive function $g(\cdot)$ if $\lim_{x\rightarrow
x_{0}}f(x)/g(x)=0$.

\subsection{Spectrally negative L\'evy processes and scale functions}

Consider a spectrally negative L\'{e}vy process $X=\{X_{t},t\geq0\}$.
Throughout the paper, we assume that $|X|$ is not a subordinator and
hence $0$
is regular for $(0,\infty)$ (see Definition~6.4 and Theorem~6.5 of Kyprianou
\cite{K06} for the definition and equivalent characterizations of the
regularity). The Laplace exponent of $X$ is given by
%
\begin{equation}
\psi(s):=\frac{1}{t}\log\mathbb{E} \bigl[e^{sX_{t}} \bigr]=-\mu s+
\frac
{1}{2}\sigma ^{2}s^{2}+\int_{(-\infty,0)}
\bigl( e^{sx}-1-sx1_{\{x>-1\}} \bigr) \Pi(\mathrm{d}x),
\label{decomp}%
\end{equation}
for every $s\in\mathbb{H}^{+}$. Here, $\sigma\geq0$ and the L\'
{e}vy measure
$\Pi(\mathrm{d}x)$ is supported on $(-\infty,0)$ with
\[
\int_{(-\infty,0)} \bigl(1\wedge x^{2} \bigr)\Pi(
\mathrm{d}x)<\infty.
\]
It is known that $X$ has paths of bounded variation if and only if
$\int_{(-1,0)}|x|\Pi(\mathrm{d}x)<\infty$ and $\sigma=0$. In this
case, we can
rewrite (\ref{decomp}) as
%
\begin{equation}
\psi(s)=sd+\int_{(-\infty,0)} \bigl(e^{sx}-1 \bigr)\Pi(
\mathrm{d}x), \qquad s\geq0, \label{psi bv}%
\end{equation}
where the drift $d:=-\mu+\int_{(-1,0)}|x|\Pi(\mathrm{d}x)>0$ as
$|X|$ is not a
subordinator. For any given $q\geq0$, the equation $\psi(s)=q$ has at least
one positive solution, and we denote the largest one by $\Phi(q)$.

It is well known that $ \{ e^{cX_{t}-\psi(c)t},t\geq0 \} $
is a
martingale for any $c\geq0$. This gives rise to the change of measure
%
\begin{equation}
\frac{\mathrm{d}\mathbb{P}^{c}}{\mathrm{d}\mathbb
{P}}\Big\vert _{\mathcal{F}_{t}}=e^{cX_{t}-\psi(c)t},\qquad t\geq0.
\label{Ps}%
\end{equation}
Under the new measure $\mathbb{P}^{c}$, $X$ is still a spectrally negative
L\'{e}vy process, and its Laplace exponent is given by $\psi_{c}%
(s)=\psi(s+c)-\psi(c)$ for all $s\in\mathbb{C}$ such that $s+c\in
\mathbb{H}^{+}$.

For any $q\geq0$, the $q$-scale function $W^{(q)}:%
\mathbb{R}
\mapsto{}[0,\infty)$ is the unique function supported on
$(0,\infty)$ with
Laplace transform
\[
\int_{(0,\infty)}e^{-sx}W^{(q)}(x)\,\mathrm{d}x=
\frac{1}{\psi(s)-q},\qquad  s>\Phi(q).
\]
It is known that $W^{(q)}$ is continuous and increasing on $(0,\infty)$.
Henceforth, we assume that the jump measure $\Pi(\mathrm{d}x)$ has no atom,
then it follows that $W^{(q)}\in C^{1}(0,\infty)$ (e.g., Lemma~2.4 of
Kuznetsov \textit{et al.} \cite{KKR12}). Moreover, if the Gaussian coefficient
$\sigma>0$ then $W^{(q)}\in C^{2}(0,\infty)$ for all $q\geq0$ (e.g., Theorem~3.10 of Kuznetsov \textit{et al.} \cite{KKR12}). The $q$-scale function
$W^{(q)}$ is
closely related to exit problems of the spectrally negative L\'{e}vy process
$X$ with respect to first passage times of the form
\[
T_{x}^{+(-)}=\inf \bigl\{ t\geq0:X_{t}\geq(\leq)x
\bigr\},\qquad x\in%
\mathbb{R} 
.
\]
Two well-known fluctuation identities of spectrally negative L\'{e}vy
processes are given below (e.g., Kyprianou \cite{K06}, Theorem~8.1). For
$q\geq0$ and $0\leq x\leq a$, we have
%
\begin{equation}
\mathbb{E}_{x} \bigl[ e^{-qT_{a}^{+}}1_{ \{
T_{a}^{+}<T_{0}^{-} \}
} \bigr] =
\frac{W^{(q)}(x)}{W^{(q)}(a)} \label{up}%
\end{equation}
and
%
\begin{equation}
\mathbb{E}_{x} \bigl[e^{-qT_{0}^{-}}1_{\{T_{0}^{-}<T_{a}^{+}\}}
\bigr]=Z^{(q)}%
(x)-Z^{(q)}(a)\frac{W^{(q)}(x)}{W^{(q)}(a)},
\label{down}%
\end{equation}
where $Z^{(q)}(x)=1+q\int_{0}^{x}W^{(q)}(y)\,\mathrm{d}y$.

The following lemma gives the behavior of scale functions at $0+$ and
$\infty
$; see, for example, Lemmas 3.1, 3.2 of Kuznetsov \textit{et al.} \cite{KKR12}. Relation
(\ref{WW}) is from (3.13) of Egami \textit{et al.} \cite{ELY13}.

\begin{lemma}
\label{lem W}For any $q\geq0$,
\begin{eqnarray*}
W^{(q)}(0+) & =& \cases{0, &\quad
$\mbox{if }\sigma>0\mbox{ or }\int_{(-1,0)}|x|\Pi(
\mathrm{d}%
x)=\infty\mbox{ (unbounded variation)},$
\vspace*{2pt}\cr
\displaystyle\frac{1}{d}, & \quad $\mbox{otherwise (bounded variation),}$}
\\
W^{(q)^{\prime}}(0+) & =& \cases{ %
\displaystyle\frac{2}{\sigma^{2}}, & \quad $\mbox{if }\sigma>0,$
\vspace*{2pt}\cr
\infty, & \quad $\mbox{if }\sigma=0\mbox{ and }\Pi(-\infty,0)=\infty,$
\vspace*{2pt}\cr
\displaystyle\frac{q+\Pi(-\infty,0)}{d^{2}}, & \quad $\mbox{if }\sigma=0\mbox{ and }\Pi (-\infty,0)<\infty,$
}
\end{eqnarray*}
and
%
\begin{equation}
\lim_{x\rightarrow\infty}\frac{W^{(q)\prime}(x)}{W^{(q)}(x)}=\Phi(q). \label{WW}%
\end{equation}
\end{lemma}

\subsection{The ascending ladder process of general L\'evy processes}

In this subsection, we consider a general L\'{e}vy process $X=\{X_{t}%
,t\geq0\}$ characterized by its characteristic exponent
%
\begin{equation}
\Psi(s):=-\frac{1}{t}\log\mathbb{E} \bigl[e^{{i}sX_{t}} \bigr]=
{i}\mu s+\frac{1}{2}\sigma^{2}s^{2}+\int
_{\mathbb{R}\setminus\{0\}
} \bigl(1-e^{{i}%
sx}+{i}sx1_{\{|x|<1\}}
\bigr)\Pi(\mathrm{d}x), \label{cha}%
\end{equation}
for all $s\in\mathbb{R}$. If $X$ has bounded variation, we can rewrite
(\ref{cha}) as
%
\begin{equation}
\Psi(s)=-{i}sd+\int_{\mathbb{R}\setminus\{0\}} \bigl(1-e^{
{i}sx}%
 \bigr)\Pi(\mathrm{d}x), \label{bv cha}%
\end{equation}
where the drift $d:=-\mu-\int_{0<|x|<1}x\Pi(\mathrm{d}x)$.

The local time of $X$ at its running maximum, denoted by $L=\{
L_{t},t\geq0\}$,
is a continuous, non-decreasing, $\mathbb{R}_{+}$-valued process. The inverse
local time process, also known as the ascending ladder time process, is
defined as $L^{-1}=\{L_{t}^{-1},t\geq0\}$ where
\[
L_{t}^{-1}:= \cases{ %
\inf\{s>0:L_{s}>t\}, & \quad$\mbox{if }t<L_{\infty},$
\vspace*{2pt}\cr
\infty, &\quad  $\mbox{otherwise.}$}
\]
The ladder height process $H=\{H_{t},t\geq0\}$ is defined as
\[
H_{t}:= \cases{ %
X_{L_{t}^{-1}},
& \quad $\mbox{if }t<L_{\infty},$
\vspace*{2pt}\cr
\infty, &\quad $\mbox{otherwise.}$}
\]
The inverse local time $L^{-1}$ corresponds to the real times at which new
maxima are reached, and the ascending ladder height process $H$
corresponds to
the set of new maxima.

The bivariate process $(L^{-1},H)=\{(L_{t}^{-1},H_{t}),t\geq0\}$,
called the
ascending ladder process of~$X$, is a two-dimensional (possibly killed)
subordinator with joint Laplace transform
\[
\mathbb{E} \bigl[e^{-\alpha L_{t}^{-1}-\beta H_{t}}1_{\{t<L_{\infty}\}}%
 \bigr]=e^{-\kappa(\alpha,\beta)t}, \qquad \alpha,\beta\geq0.
\]
The joint Laplace exponent is given by
%
\begin{equation}
\kappa(\alpha,\beta)=\kappa(0,0)+\alpha d_{L}+\beta
d_{H}+\int_{(0,\infty
)^{2}} \bigl(1-e^{-\alpha x-\beta y} \bigr)
\Lambda(\mathrm{d}x,\mathrm{d}y), \qquad \alpha,\beta\geq0, \label{kappa}%
\end{equation}
where $(d_{L},d_{H})\in%
\mathbb{R}
_{+}^{2}$ and $\Lambda$ is a bivariate intensity measure on $(0,\infty)^{2}$
satisfying
\[
\int_{(0,\infty)^{2}} \bigl(1\wedge\sqrt{x^{2}+y^{2}}
\bigr)\Lambda(\mathrm{d} 
x,\mathrm{d}y)<\infty.
\]
When $L^{-1}$ and $H$ are independent, $\Lambda$ takes the form
$\Lambda
(\mathrm{d}x,\mathrm{d}y)=\Lambda_{L}(\mathrm{d}x)\delta
_{0}(\mathrm{d}%
y)+ \Lambda_{H}(\mathrm{d}y)\delta_{0}(\mathrm{d}x)$ for $x,y\geq
0$. In
particular, if $X$ is a spectrally negative L\'{e}vy process, one can choose
$L_{t}=M_{t}$, which implies that $L_{t}^{-1}=T_{t}^{+}$,
$H_{t}=X_{T_{t}^{+}%
}=t$ on $\{t<L_{\infty}\}$, and further $\kappa(\alpha,\beta)=\Phi
(\alpha)+\beta$.

By letting $\beta=0$ in (\ref{kappa}), we obtain the Laplace exponent
of the
ascending ladder time process,
\[
-\frac{1}{t}\log\mathbb{E} \bigl[e^{-\alpha L_{t}^{-1}}1_{\{t<L_{\infty}\}
}%
 \bigr]=\kappa(\alpha,0)=\kappa(0,0)+\alpha d_{L}+\int
_{(0,\infty
)} \bigl(1-e^{-\alpha
x} \bigr)\nu_{L}(
\mathrm{d}x), \qquad\alpha\geq0,
\]
where $\nu_{L}(\mathrm{d}x)=\Lambda(\mathrm{d}x,(0,\infty))$ is
the jump
measure of $L^{-1}$. It follows from integration by parts that
%
\begin{equation}
\kappa(\alpha,0)-\kappa(0,0)=\alpha \biggl( d_{L}+\int
_{(0,\infty
)}e^{-\alpha
x}\bar{\nu}_{L}(x)\,\mathrm{d}x
\biggr) ,\qquad  \alpha\geq0, \label{nuT}%
\end{equation}
where $\bar{\nu}_{L}(x):=\nu_{L}(x,\infty)$.

The renewal function $h$ associated with the ladder height process $H$ is
defined as
%
\begin{equation}
h(x):=\int_{(0,\infty)}\mathbb{P}\{H_{t}\leq x\}
\,\mathrm{d}t,\qquad  x\geq0.\label{renewal}%
\end{equation}
When $X$ is a spectrally negative L\'{e}vy process, it is easily seen that
$h(x)=\int_{(0,x)}e^{-\Phi(0)t}\,\mathrm{d}t$ for $x\geq0$. We recall
the follow
results (see Theorem~5 in Chapter III and Theorem~19 in Chapter VI of Bertoin
\cite{B96}) on the connection between the renewal function and the creeping
property. Here we say $X$ creeps across $x$ if it enters $(x,\infty)$
continuously.

\begin{lemma}
\label{lem H}The following assertions are equivalent.
\begin{longlist}[(iii)]
\item[(i)] $\mathbb{P}\{X$ creeps across $x\}>0$ for some $x>0$.

\item[(ii)] The drift coefficient $d_{H}>0$.

\item[(iii)] The renewal function $h$ is absolute continuous and
$h^{\prime}$
is bounded.
\end{longlist}

Moreover, when these assertions hold, there is a version $h^{\prime}$
that is
continuous and positive on $(0,\infty)$. Finally, $\lim_{x\downarrow
0}h^{\prime}(x)=\frac{1}{d_{H}}>0$ and $\mathbb{P}\{X$ creeps across
$x\}=d_{H}h^{\prime}(x)$ for all $x>0$.
\end{lemma}

\section{Magnitude of drawdowns revisited}
\label{sec:mag}

In this section, we revisit some known results of the magnitude of drawdowns
of spectrally negative L\'{e}vy processes via fluctuation identities
and an
approximation approach introduced by Lehoczky \cite{L77}. Such
approach is in
the spirit of the general It\^{o}'s excursion theory.

\begin{lemma}
For $q\geq0$ and $x>0$, we have
%
\begin{equation}
\mathbb{E} \bigl[ e^{-qT_{x}^{+}}1_{ \{ M_{\tau_{a}}\geq x
\}
} \bigr] =\exp \biggl\{ -
\frac{W^{(q)\prime}(a)}{W^{(q)}(a)}x \biggr\} . \label{TM}%
\end{equation}
\end{lemma}

\begin{pf}
For fixed $x>0$ and $n\in%
\mathbb{N}
$, let $\{s_{n,i},i=0,\ldots,n\}$ be a sequence of increasing
partitions of
the interval $[0,x]$ with $0=s_{n,0}<s_{n,1}<\cdots<s_{n,n}=x$ and
such that
$\Delta_{n}=\max_{1\leq i\leq n}(s_{n,i}-s_{n,i-1})$ decreases to $0$ as
$n\rightarrow\infty$. Using the strong Markov property of $X$, we
propose to
approximate the event $ ( M_{\tau_{a}}\geq x ) $ by
$\bigcap_{m=1}^{n}(  T_{s_{n,i}}^{+}<T_{s_{n,i-1}-a}^{-}\vert
X_{0}=s_{n,i-1} ) $, and thus use
\[
E_{n}:=\prod_{i=1}^{n}
\mathbb{E} \bigl[ e^{-qT_{s_{n,i}}^{+}%
}1_{ \{ T_{s_{n,i}}^{+}<T_{s_{n,i-1}-a}^{-} \} }\vert X_{0}=s_{n,i-1}
\bigr] ,
\]
as an approximation of $\mathbb{E} [ e^{-qT_{x}^{+}}1_{ \{
M_{\tau_{a}}\geq x \} } ] $. By (\ref{up}), we have
\[
E_{n}=\prod_{i=1}^{n}
\frac{W^{(q)}(a)}{W^{(q)}(a+s_{n,i}-s_{n,i-1})}%
=\exp \Biggl\{ \sum_{i=1}^{n}
\ln \biggl\{ 1-\frac{W^{(q)}(a+s_{n,i}%
-s_{n,i-1})-W^{(q)}(a)}{W^{(q)}(a+s_{n,i}-s_{n,i-1})} \biggr\} \Biggr\}.
\]
Since $W^{ ( q ) }\in C^{1}(0,\infty)$ and is increasing on
$(0,\infty)$, we have
\[
\biggl( \frac
{W^{(q)}(a+s_{n,i}-s_{n,i-1})-W^{(q)}(a)}{W^{(q)}(a+s_{n,i}%
-s_{n,i-1})} \biggr) ^{2}\leq \biggl( \frac{W^{(q)}(a+\Delta
_{n})-W^{(q)}%
(a)}{W^{(q)}(a)}
\biggr) ^{2}\leq K(\Delta_{n})^{2},%
\]
for all $1\leq i\leq n$ and some constant $K>0$. By the fact that
$-\ln(1-\varepsilon)=\varepsilon+o(\varepsilon)$ for small
$\varepsilon>0$, it
follows that
\begin{eqnarray*}
&&\mathbb{E} \bigl[ e^{-qT_{x}^{+}}1_{ \{ M_{\tau_{a}}\geq x
\}
} \bigr] \\
&&\quad =\lim
_{n\rightarrow\infty}\exp \Biggl\{ \sum_{i=1}^{n}
\ln \biggl\{ 1-\frac
{W^{(q)}(a+s_{n,i}-s_{n,i-1})-W^{(q)}(a)}{W^{(q)}(a+s_{n,i}-s_{n,i-1}%
)} \biggr\} \Biggr\}
\\
&&\quad =\lim_{n\rightarrow\infty}\exp \Biggl\{ -\sum
_{i=1}^{n}\frac
{W^{(q)}%
(a+s_{n,i}-s_{n,i-1})-W^{(q)}(a)}{W^{(q)}(a+s_{n,i}-s_{n,i-1})} \Biggr\}
\\
&&\quad =\exp \biggl\{ -\frac{W^{(q)\prime}(a)}{W^{(q)}(a)}x \biggr\} ,
\end{eqnarray*}
which completes the proof.
\end{pf}

By letting $q=0$ in (\ref{TM}), it is easy to see that $M_{\tau_{a}}$ follows
an exponential distribution with mean $W(a)/W^{\prime}(a)$. Then it follows
from (\ref{TM}) that, for $q\geq0$ and $x\geq0$,
%
\begin{equation}
\mathbb{E} \bigl[ e^{-qT_{x}^{+}}\vert M_{\tau
_{a}}=x \bigr] = \exp \biggl
\{ - \biggl( \frac{W^{(q)\prime}(a)}{W^{(q)}(a)}-\frac
{W^{\prime
}(a)}{W(a)} \biggr) x \biggr\}.
\label{TM0}%
\end{equation}

Next, we consider the following lemma which relates to downward exiting.

\begin{lemma}
For $q,s\geq0$, we have
%
\begin{equation}
\mathbb{E}_{a} \bigl[ e^{-qT_{0}^{-}-s(a-X_{T_{0}^{-}})}\vert T_{0}^{-}<T_{a}^{+}
\bigr] = \frac{W(a)}{W^{\prime}(a)}\frac{Z_{s}%
^{(p)}(a)W_{s}^{(p)\prime}(a)-pW_{s}^{(p)}(a)^{2}}{W_{s}^{(p)}(a)}, \label{ricochet}%
\end{equation}
where $p=q-\psi(s)$, $W_{s}^{(p)}$ and $Z_{s}^{(p)}$ are $p$-scale functions
under $\mathbb{P}^{s}$.
\end{lemma}

\begin{pf}
We first consider that $s\leq\Phi(q)$, or equivalently, $q\geq\psi
(s)$. For
$0\leq x\leq y$, since $T_{0}^{-}\wedge T_{y}^{+}$ is a.s. finite, by change
of measure (\ref{Ps}) and (\ref{down}),
%
\begin{eqnarray} \label{down LT}%
\mathbb{E}_{x} \bigl[ e^{-qT_{0}^{-}-s(x-X_{T_{0}^{-}})}1_{\{T_{0}^{-}%
<T_{y}^{+}\}} \bigr]& =&
\mathbb{E}_{x}^{s} \bigl[e^{-pT_{0}^{-}}1_{\{T_{0}^{-}
<T_{y}^{+}\}}
\bigr]
\nonumber
\\[-8pt]
\\[-8pt]
\nonumber
&=&Z_{s}^{(p)}(x)-Z_{s}^{(p)}(y)
\frac{W_{s}^{(p)}(x)}{W_{s}
^{(p)}(y)}.
\end{eqnarray}
It follows from (\ref{up}) and (\ref{down LT}) that
%
\begin{eqnarray}\label{result}%
&&\mathbb{E}_{a} \bigl[ e^{-qT_{0}^{-}-s(a-X_{T_{0}^{-}})}|T_{0}^{-}<T_{a}%
^{+}
\bigr]\nonumber\\
 &&\quad =\lim_{\varepsilon\downarrow0}\mathbb{E}_{a} \bigl[
e^{-qT_{0}^{-}-s(a-X_{T_{0}^{-}})}|T_{0}^{-}<T_{a+\varepsilon
}^{+}
\bigr]
\nonumber
\\[-8pt]
\\[-8pt]
\nonumber
&&\quad =\lim_{\varepsilon\downarrow0} \biggl( Z_{s}^{(p)}(a)-Z_{s}^{(p)}%
(a+
\varepsilon)\frac{W_{s}^{(p)}(a)}{W_{s}^{(p)}(a+\varepsilon
)} \biggr) \frac{W(a+\varepsilon)}{W(a+\varepsilon)-W(a)}
\\
&&\quad =\frac{W(a)}{W^{\prime}(a)}\frac{Z_{s}^{(p)}(a)W_{s}^{(p)\prime}%
(a)-pW_{s}^{(p)}(a)^{2}}{W_{s}^{(p)}(a)}. \nonumber
\end{eqnarray}
The other side of the approximation $\lim_{\varepsilon\downarrow0}%
\mathbb{E}_{a-\varepsilon} [
e^{-qT_{0}^{-}-s(a-X_{T_{0}^{-}})}|T_{0}%
^{-}<T_{a}^{+} ] $ also results in~(\ref{result}). The proof is then
completed through an analytical extension of (\ref{ricochet}) to
$s\geq
0$.
\end{pf}

To obtain the main result of this section, we notice that a sample path
of $X$
until $\tau_{a}$ can be splitted into two parts: the rising part and the
subsequent crashing part. Because of the regularity of $0$ for
$(0,\infty)$,
we know that the last passage time $(G_{\tau_{a}}|M_{\tau
_{a}}=x)=(T_{x}%
^{+}|M_{\tau_{a}}=x)$, $\mathbb{P}$-a.s. (see also discussions on
page 158 of
Kyprianou \cite{K06}). Our analysis essentially follows this idea: relations
(\ref{TM0}) and (\ref{ricochet}) correspond to the rising and the crashing
part, respectively. The following quadruple LT is obtained by pasting these
two parts at the turning point $G_{\tau_{a}}$.

\begin{theorem}
For $q,r,s,\delta\geq0$, we have
%
\begin{equation}
\mathbb{E} \bigl[e^{-q\tau_{a}-rG_{\tau_{a}}-sY_{\tau_{a}}-\delta M_{\tau
_{a}}%
} \bigr]=\frac{W^{(q+r)}(a)}{\delta W^{(q+r)}(a)+W^{(q+r)\prime}(a)}\frac
{Z_{s}^{(p)}(a)W_{s}^{(p)\prime}(a)-pW_{s}^{(p)}(a)^{2}}{W_{s}^{(p)}(a)},
\label{4LT}%
\end{equation}
where $p=q-\psi(s)$.
\end{theorem}

\begin{pf}
By conditioning on the event $(M_{\tau_{a}}=x)$ for $x>0$, we have
$\tau
_{a}=G_{\tau_{a}}+T_{x-a}^{-}\circ\theta_{G_{\tau_{a}}}$ and
$T_{x-a}^{-}%
\circ\theta_{G_{\tau_{a}}}<T_{x}^{+}\circ\theta_{G_{\tau_{a}}}$,
$\mathbb{P}%
$-a.s. where $\theta$ is the Markov shift operator defined as
$X_{t}\circ
\theta_{s}=X_{t+s}$. Therefore, by (\ref{TM0}) and (\ref{ricochet}), we
obtain
%
\begin{eqnarray}\label{last}%
&& \mathbb{E} \bigl[e^{-q\tau_{a}-rG_{\tau_{a}}-sY_{\tau_{a}}}|M_{\tau_{a}
}=x \bigr]
\nonumber
\\
&&\quad =\mathbb{E} \bigl[ e^{-(q+r)G_{\tau_{a}}-q(\tau_{a}-G_{\tau_{a}}%
)-sY_{\tau_{a}}}|M_{\tau_{a}}=x \bigr]
\nonumber
\\
& &\quad=\mathbb{E} \bigl[ e^{-(q+r)G_{\tau_{a}}}\mathbb{E} \bigl[ e^{-qT_{x-a}^{-}\circ\theta_{G_{\tau
_{a}}}-s(x-X_{T_{x-a}^{-}})}\vert
T_{x-a}^{-}\circ \theta_{G_{\tau_{a}}}<T_{x}^{+}
\circ\theta _{G_{\tau_{a}}%
} \bigr] \vert M_{\tau_{a}}=x \bigr]
\\
&&\quad =\mathbb{E} \bigl[e^{-(q+r)G_{\tau_{a}}}|M_{\tau_{a}}=x \bigr]\mathbb
{E}_{x} \bigl[ e^{-qT_{x-a}^{-}-s(x-X_{T_{x-a}^{-}})}\vert T_{x-a}^{-}%
<T_{x}^{+}
\bigr]
\nonumber
\\
&&\quad =\exp \biggl\{ - \biggl( \frac{W^{(q+r)\prime}(a)}{W^{(q+r)}(a)}%
-\frac{W^{\prime}(a)}{W(a)}
\biggr) x \biggr\} \frac{W(a)}{W^{\prime
}(a)}%
\frac{Z_{s}^{(p)}(a)W_{s}^{(p)\prime}(a)-pW_{s}^{(p)}(a)^{2}}{W_{s}^{(p)}(a)}.
\nonumber
\end{eqnarray}
Multiplying (\ref{last}) by the density of $M_{\tau_{a}}$ and then integrating
with respect to $x$, we obtain~(\ref{4LT}).
\end{pf}

Relation (\ref{4LT}) generalizes Theorem~1 of Avram \textit{et al.} \cite
{AAP04} by
incorporating the joint LT of $G_{\tau_{a}}$ and $M_{\tau_{a}}$.
Moreover, by
a similar approximation argument, one can solve for the joint
distribution of
equation (\ref{4LT}) but with the law of $Y_{\tau_{a}}$, which then recovers
the sextuple law in Theorem~1 of Mijatovi\'{c} and Pistorius \cite{MP12} (the
running minimum at $\tau_{a}$ can also be easily incorporated).

\section{Asymptotics of magnitude of drawdowns}
\label{sec:asym}

In this section, we investigate the asymptotics of the LT (\ref{4LT})
of the
magnitude of drawdowns as $a\downarrow0$ for spectrally negative L\'{e}vy
processes. Furthermore, we show that such asymptotics are robust with
respect to the perturbation by positive compound Poisson jumps.

\subsection{Spectrally negative L\'evy processes}

Henceforth, we make the following assumption on the behavior of the scale
function at $0+$.

\begin{assumption}
\label{assume1}
\[
\lim_{x\downarrow0}xW^{\prime}(x)=0.
\]
\end{assumption}

In fact, since $xW^{\prime}(x)\geq0$ for all $x>0$, as long as
$W^{\prime}$ is
well-behaved at $0+$ in the sense that
\[
\lim_{x\downarrow0}xW^{\prime}(x)=c\qquad\mbox{for some }c\in{}[ 0,
\infty],
\]
one deduces from the integrability of $W^{\prime}$ at $0+$ that $c=0$.

\begin{remark}
\label{rk assume1}From Lemma~\ref{lem W}, it is clear that Assumption~\ref{assume1} holds if the Gaussian component $\sigma>0$ or $\Pi
(-\infty,0)<\infty$. Moreover, the spectrally negative $\alpha
$-stable process
with index $\alpha\in(1,2)$, whose Laplace exponent $\psi
(s)=s^{\alpha}$ and
scale function
\[
W(x)=1_{\{x\geq0\}}\frac{x^{\alpha-1}}{\Gamma(\alpha)},
\]
also satisfies Assumption~\ref{assume1}.
\end{remark}

Since scale functions are only known in a few cases, we examine sufficient
conditions on the Laplace exponent to identify cases when Assumption~\ref{assume1} holds.

\begin{remark}
\label{rk char}For a general spectrally negative L\'{e}vy process with Laplace
exponent $\psi$, by Lemma~\ref{lem W}, one can choose an arbitrary
$s_{0}%
>\Phi(0)$ and define a function $g(x):=1_{\{x>0\}
}e^{-s_{0}x}xW^{\prime}(x)$,
which is non-negative and continuous on $\mathbb{R}\setminus\{0\}$.
By Lemma~3.3 of Kuznetsov \cite{KKR12} and (\ref{WW}), we further know that
$g(x)\in
L^{1}(\mathbb{R})$. By integration by parts and analytical
continuation, one
obtains that
\[
\int_{\mathbb{R}}e^{{i}sx}g(x)\,\mathrm{d}x=\varphi
(s_{0}-{i}%
s),\qquad  s\in\mathbb{R},
\]
where $\varphi(s):=\frac{s\psi^{\prime}(s)-\psi(s)}{\psi(s)^{2}}$ for
$\operatorname{Re}(s)\geq0$. By the Fourier inversion and the dominated
convergence theorem, we know that a sufficient condition for Assumption~\ref{assume1} to hold is that $\varphi(s_{0}-{i}\cdot)\in
L^{1}(\mathbb{R})$ as it implies that $g(\cdot)$ is continuous over
$\mathbb{R}$.
\end{remark}

\begin{lemma}
\label{lem asp 1}Under Assumption~\ref{assume1}, we have $\lim_{x\downarrow
0}xW^{(q)\prime}(x)=0$ for every $q\geq0$.
\end{lemma}

\begin{pf}
Since the scale function $W$ is supported on $(0,\infty)$, for any
$k\geq1$,
we have
%
\begin{eqnarray} \label{dk}%
\frac{\mathrm{d}}{\mathrm{d}x}W^{\ast(k+1)}(x) & =&\int_{(0,x)}W^{\prime
}(x-y)W^{\ast k}(y)
\,\mathrm{d}y+W(0+)W^{\ast k}(x)
\nonumber
\\
& \leq&\frac{x^{k-1}}{(k-1)!}W^{k}(x) \biggl( \int_{(0,x)}W^{\prime
}(x-y)
\,\mathrm{d}y+W(0+) \biggr)
\\
& =&\frac{x^{k-1}}{(k-1)!}W^{k+1}(x),\nonumber
\end{eqnarray}
where the inequality above is due to equation (8.23) of Kyprianou \cite{K06}
and the monotonicity of $W$. By (\ref{dk}) and taking derivatives term
by term
to the well-known identity $W^{(q)}(x)=\sum_{k=0}^{\infty
}q^{k}W^{\ast
(k+1)}(x)$, where $W^{\ast k}$ is the $k$th convolution of $W$ with itself,
we obtain
\begin{eqnarray*}
xW^{(q)\prime}(x) & =&xW^{\prime}(x)+x\sum_{k=1}^{\infty}q^{k}
\frac
{\mathrm{d}}{\mathrm{d}x}W^{\ast(k+1)}(x)
\\
& \leq& xW^{\prime}(x)+qxW^{2}(x)\sum
_{k=1}^{\infty}\frac{ (
qxW(x) ) ^{k-1}}{(k-1)!}
\\
& =&xW^{\prime}(x)+qxW^{2}(x)e^{qxW(x)}.
\end{eqnarray*}
This ends the proof as the right-hand side of the last equation
approaches $0$
as $x\downarrow0$ by Assumption~\ref{assume1}.
\end{pf}

Lemma~\ref{lem asp 1} is paramount to derive the following asymptotic results.

\begin{theorem}
\label{thm asym}Consider a spectrally negative L\'{e}vy process $X$ satisfying
Assumption~\ref{assume1}. For any $q,s\geq0$, we have
\[
\lim_{\varepsilon\downarrow0}\frac{W^{(q)\prime}(\varepsilon
)}{W^{(q)}%
(\varepsilon)} \bigl(1-\mathbb{E}
\bigl[e^{-q\tau_{\varepsilon}-sY_{\tau
_{\varepsilon}}%
} \bigr] \bigr)= \cases{s, &\quad $\mbox{if }X\mbox{ has unbounded variation,}$
\vspace*{2pt}\cr
\displaystyle s+\frac{q-\psi(s)}{d}, &\quad $\mbox{if }X\mbox{ has bounded variation.}$}
\]
\end{theorem}

\begin{pf}
Using (\ref{4LT}), one deduces that
%
\begin{eqnarray}\label{2LT}%
&&\frac{W^{(q)\prime}(\varepsilon)}{W^{(q)}(\varepsilon)} \bigl( 1-\mathbb{E}%
 \bigl[e^{-q\tau_{\varepsilon}-sY_{\tau_{\varepsilon}}} \bigr]
\bigr)\nonumber \\
&&\quad =s- \bigl(q-\psi (s) \bigr)\frac{W^{(q)\prime}(\varepsilon)}{W^{(q)}(\varepsilon)}\int_{(0,\varepsilon)}e^{-sx}W^{(q)}(x)
\,\mathrm{d}x
\\
& &\qquad{}+s \bigl(q-\psi(s) \bigr)\int_{(0,\varepsilon)}e^{-sx}W^{(q)}(x)
\,\mathrm{d}%
x+ \bigl(q-\psi(s) \bigr)e^{-s\varepsilon}W^{(q)}(
\varepsilon). \nonumber
\end{eqnarray}
From the monotonicity of $W^{(q)}(\cdot)$, we have
\[
0\leq\frac{W^{(q)\prime}(\varepsilon)}{W^{(q)}(\varepsilon)}\int_{(0,\varepsilon)}e^{-sx}W^{(q)}(x)
\,\mathrm{d}x\leq\frac
{W^{(q)\prime
}(\varepsilon)}{W^{(q)}(\varepsilon)}\varepsilon W^{(q)}(\varepsilon )=
\varepsilon W^{(q)\prime}(\varepsilon).
\]
It follows from (\ref{2LT}) and Lemma~\ref{lem asp 1} that
\[
\lim_{\varepsilon\downarrow0}\frac{W^{(q)\prime}(\varepsilon
)}{W^{(q)}%
(\varepsilon)} \bigl(1-\mathbb{E}
\bigl[e^{-q\tau_{\varepsilon}-sY_{\tau
_{\varepsilon}}%
} \bigr] \bigr)=s+ \bigl(q-\psi(s) \bigr)W^{(q)}(0+),
\]
which ends the proof by Lemma~\ref{lem W}.
\end{pf}

\subsection{A class of L\'evy models with two-sided jumps}

Next, we consider a class of L\'{e}vy process with two-sided jumps of
the form
%
\begin{equation}
X_{t}=\tilde{X}_{t}+S_{t}^{+},
\label{Xph}%
\end{equation}
where $\tilde{X}$ a spectrally negative L\'{e}vy process satisfying Assumption~\ref{assume1}, and $S^{+}$ is a compound Poisson process with arrival rate
$\lambda^{+}=\Pi(0,\infty)\in(0,\infty)$ and i.i.d. positive jump
size with
distribution function $F^{+}$. The two processes $\tilde{X}$ and
$S^{+}$ are
assumed to be independent. Since we assume that $|\tilde{X}|$ is not a
subordinator and is regular for $(0,\infty)$, it is clear that the
same holds
for~$X$.

The characteristic exponent of $X$ is given by
%
\begin{equation}
\Psi(s)=\tilde{\Psi}(s)+\lambda^{+}\int_{0}^{\infty}
\bigl( 1-e^{isx} \bigr) F^{+}(\mathrm{d}x),\qquad  s\in\mathbb{R},
\label{cha bv}%
\end{equation}
where $\tilde{\Psi}(\cdot)$ is the characteristic exponent of
$\tilde{X}$.
Henceforth, we add the symbol $\ \widetilde\ $ \ to all quantities
when they
relate to the spectrally negative L\'{e}vy component $\tilde{X}$ only.

By conditioning on the first positive jump arrival time and the jump
size, we
have the following representation of the joint Laplace transform of
$(\tau_{\varepsilon},Y_{\tau_{\varepsilon}})$.

\begin{lemma}
For $q,s\geq0$ and $\varepsilon>0$, we have
%
\begin{equation}
\mathbb{E} \bigl[e^{-q\tau_{\varepsilon}-sY_{\tau_{\varepsilon}}} \bigr]=\frac
{\mathbb{E}[e^{- ( q+\lambda^{+} ) \tilde{\tau
}_{\varepsilon
}-s\tilde{Y}_{\tilde{\tau}_{\varepsilon}}}]+\mathbb{E}[e^{-q\tau
_{\varepsilon
}-sY_{\tau_{\varepsilon}}}1_{\{\tau_{\varepsilon}>\xi
_{1}^{+},J_{1}^{+}%
<Y_{\xi_{1}^{+}-}\}}]}{1-({\lambda^{+}}/{(q+\lambda^{+})}) (
1-\mathbb{E}[e^{-(q+\lambda^{+})\tilde{\tau}_{\varepsilon}}] )
+\mathbb{E}[e^{-q\xi_{1}^{+}}1_{\{\tilde{\tau}_{\varepsilon}>\xi
_{1}^{+}%
,J_{1}^{+}<\tilde{Y}_{\xi_{1}^{+}-}\}}]},
\label{21}%
\end{equation}
where $\xi_{1}^{+}$ and $J_{1}^{+}$ are the time and size of the first upward
jump of $X$, respectively.
\end{lemma}

\begin{pf}
Recall that $\xi_{1}^{+}$ is exponentially distributed with mean
$1/\lambda^{+}$. By the strong Markov property of $X$ and the fact that
$(\tau_{\varepsilon}<\xi_{1}^{+})=(\tilde{\tau}_{\varepsilon}<\xi
_{1}^{+})$
a.s.,
\begin{eqnarray*}
&& \mathbb{E} \bigl[e^{-q\tau_{\varepsilon}-sY_{\tau_{\varepsilon}}} \bigr]
\\
&&\quad =\mathbb{E} \bigl[e^{-q\tau_{\varepsilon}-sY_{\tau_{\varepsilon}}}1_{\{
\tau
_{\varepsilon}<\xi_{1}^{+}\}} \bigr]+\mathbb{E}
\bigl[e^{-q\tau_{\varepsilon}%
-sY_{\tau_{\varepsilon}}}1_{\{\tau_{\varepsilon}>\xi_{1}^{+}\}} \bigr]
\\
&&\quad =\mathbb{E} \bigl[e^{-q\tilde{\tau}_{\varepsilon}-s\tilde{Y}_{\tilde
{\tau
}_{\varepsilon}}}1_{\{\tilde{\tau}_{\varepsilon}<\xi_{1}^{+}\}
} \bigr]+
\mathbb{E}%
 \bigl[e^{-q\tau_{\varepsilon}-sY_{\tau_{\varepsilon}}}1_{\{\tau
_{\varepsilon}%
>\xi_{1}^{+},J_{1}^{+}\geq Y_{\xi_{1}^{+}-}\}} \bigr]\\
&&\qquad{}+\mathbb{E}
\bigl[e^{-q\tau
_{\varepsilon}-sY_{\tau_{\varepsilon}}}1_{\{\tau_{\varepsilon}>\xi
_{1}%
^{+},J_{1}^{+}<Y_{\xi_{1}^{+}-}\}} \bigr]
\\
&&\quad =\mathbb{E} \bigl[e^{- ( q+\lambda^{+} ) \tilde{\tau
}_{\varepsilon
}-s\tilde{Y}_{\tilde{\tau}_{\varepsilon}}} \bigr]+\mathbb{E} \bigl[e^{-q\xi
_{1}^{+}%
}1_{\{\tilde{\tau}_{\varepsilon}>\xi_{1}^{+},J_{1}^{+}\geq\tilde
{Y}_{\xi
_{1}^{+}-}\}}
\bigr]\mathbb{E} \bigl[e^{-q\tau_{\varepsilon}-sY_{\tau
_{\varepsilon}}} \bigr]
\\
&&\qquad{} +\mathbb{E} \bigl[e^{-q\tau_{\varepsilon}-sY_{\tau_{\varepsilon}}}1_{\{
\tau
_{\varepsilon}>\xi_{1}^{+},J_{1}^{+}<Y_{\xi_{1}^{+}-}\}} \bigr].%
\end{eqnarray*}
Solving for $\mathbb{E}[e^{-q\tau_{\varepsilon}-sY_{\tau
_{\varepsilon}}}]$,
one obtains
%
\begin{equation}
\mathbb{E} \bigl[e^{-q\tau_{\varepsilon}-sY_{\tau_{\varepsilon}}} \bigr]=\frac
{\mathbb{E}[e^{- ( q+\lambda^{+} ) \tilde{\tau
}_{\varepsilon
}-s\tilde{Y}_{\tilde{\tau}_{\varepsilon}}}]+\mathbb{E}[e^{-q\tau
_{\varepsilon
}-sY_{\tau_{\varepsilon}}}1_{\{\tau_{\varepsilon}>\xi
_{1}^{+},J_{1}^{+}%
<Y_{\xi_{1}^{+}-}\}}]}{1-\mathbb{E}[e^{-q\xi_{1}^{+}}1_{\{\tilde
{\tau
}_{\varepsilon}>\xi_{1}^{+},J_{1}^{+}\geq\tilde{Y}_{\xi_{1}^{+}-}\}}]}.
\label{pli2}%
\end{equation}
For the denominator on the right-hand side of (\ref{pli2}), we notice that
%
\begin{eqnarray} \label{pli3}%
&&\mathbb{E} \bigl[e^{-q\xi_{1}^{+}}1_{\{\tilde{\tau}_{\varepsilon}>\xi
_{1}^{+}%
,J_{1}^{+}\geq\tilde{Y}_{\xi_{1}^{+}-}\}} \bigr]\nonumber\\
&&\quad =\mathbb{E}
\bigl[e^{-q\xi
_{1}^{+}%
} \bigr]-\mathbb{E} \bigl[e^{-q\xi_{1}^{+}}1_{\{\tilde{\tau}_{\varepsilon}<\xi
_{1}^{+}%
\}}
\bigr]-\mathbb{E} \bigl[e^{-q\xi_{1}^{+}}1_{\{\tilde{\tau}_{\varepsilon
}>\xi_{1}%
^{+},J_{1}^{+}<\tilde{Y}_{\xi_{1}^{+}-}\}} \bigr]
\\
&&\quad =\frac{\lambda^{+}}{q+\lambda^{+}} \bigl( 1-\mathbb {E} \bigl[e^{-(q+\lambda
^{+})\tilde{\tau}_{\varepsilon}} \bigr] \bigr)
- \mathbb{E} \bigl[e^{-q\xi
_{1}^{+}%
}1_{\{\tilde{\tau}_{\varepsilon}>\xi_{1}^{+},J_{1}^{+}<\tilde
{Y}_{\xi_{1}%
^{+}-}\}} \bigr].\nonumber
\end{eqnarray}
The proof of (\ref{21}) is completed by substituting (\ref{pli3}) to
(\ref{pli2}).
\end{pf}

We present an analogue of Theorem~\ref{thm asym} for the L\'{e}vy process
(\ref{Xph}) with two-sided jumps. Note that by (\ref{cha bv}), the
drift of
the characteristic exponent $d$ of $\tilde{X}$ and $X$ are the same when
$\tilde{X}$ has bounded variation.

\begin{theorem}
\label{thm asym2}Consider the L\'{e}vy model (\ref{Xph}). For
$q,s\geq0$, we
have
\[
\lim_{\varepsilon\downarrow0}\frac{W^{(q+\lambda^{+})\prime
}(\varepsilon
)}{W^{(q+\lambda^{+})}(\varepsilon)} \bigl(1-\mathbb{E}
\bigl[e^{-q\tau
_{\varepsilon
}-sY_{\tau_{\varepsilon}}} \bigr] \bigr)= \cases{
s, & \quad $\mbox{if }\tilde{X}\mbox{ has unbounded variation},$
\vspace*{2pt}\cr
\displaystyle s+\frac{q-\tilde{\psi}(s)}{d}, & \quad$ \mbox{if }\tilde{X}\mbox{ has bounded
variation}.$}
\]
\end{theorem}

\begin{pf}
Since $\tilde{X}$ and $S^{+}$ are independent and $(\tau_{\varepsilon
}<\xi
_{1}^{+})=(\tilde{\tau}_{\varepsilon}<\xi_{1}^{+})$ a.s., we have
%
\begin{eqnarray}\label{joint}%
\mathbb{P} \bigl\{ \tau_{\varepsilon}>\xi_{1}^{+},J_{1}^{+}<Y_{\xi
_{1}^{+}%
-}
\bigr\} & =&\mathbb{P} \bigl\{ \tilde{\tau}_{\varepsilon}>\xi
_{1}%
^{+},J_{1}^{+}<
\tilde{Y}_{\xi_{1}^{+}-} \bigr\}
\nonumber
\\
& \leq&\mathbb{P} \bigl\{\tilde{\tau}_{\varepsilon}>\xi_{1}^{+},J_{1}^{+}
<\varepsilon \bigr\}
\nonumber
\\[-8pt]
\\[-8pt]
\nonumber
& = &\bigl( 1-\mathbb{E} \bigl[e^{-\lambda^{+}\tilde{\tau}_{\varepsilon
}} \bigr] \bigr) \mathbb{P} \bigl
\{J_{1}^{+}<\varepsilon \bigr\}
\\
& \leq& \bigl( 1-\mathbb{E} \bigl[e^{-(q+\lambda^{+})\tilde{\tau
}_{\varepsilon}%
} \bigr] \bigr) F^{+}(
\varepsilon). \nonumber
\end{eqnarray}
It follows from Theorem~\ref{thm asym} that
%
\begin{eqnarray}\label{oeps}%
&&\frac{W^{(q+\lambda^{+})\prime}(\varepsilon)}{W^{(q+\lambda^{+})}%
(\varepsilon)}\mathbb{P} \bigl\{ \tau_{\varepsilon}>\xi
_{1}^{+},J_{1}%
^{+}<Y_{\xi_{1}^{+}-}
\bigr\}
\nonumber
\\[-8pt]
\\[-8pt]
\nonumber
&&\quad\leq\frac{W^{(q+\lambda^{+})\prime}%
(\varepsilon)}{W^{(q+\lambda^{+})}(\varepsilon)} \bigl( 1-\mathbb{E} 
 \bigl[e^{-(q+\lambda^{+})\tilde{\tau}_{\varepsilon}} \bigr] \bigr) F^{+}(\varepsilon )=o(1),
\end{eqnarray}
for small $\varepsilon>0$. By (\ref{21}), (\ref{joint}) and (\ref
{oeps}), one
obtains that
%
\begin{eqnarray}\label{qq} 
&& \frac{W^{(q+\lambda^{+})\prime}(\varepsilon)}{W^{(q+\lambda^{+}%
)}(\varepsilon)} \bigl(1-\mathbb{E} \bigl[e^{-q\tau_{\varepsilon}-sY_{\tau
_{\varepsilon}}%
} \bigr] \bigr)
\nonumber
\\[-8pt]
\\[-8pt]
\nonumber
&&\quad =\frac{\frac{W^{(q+\lambda^{+})\prime}(\varepsilon
)}{W^{(q+\lambda^{+}%
)}(\varepsilon)} ( 1-\mathbb{E}[e^{-(q+\lambda^{+})\tilde{\tau
}_{\varepsilon}-s\tilde{Y}_{\tilde{\tau}_{\varepsilon}}}]-\frac
{\lambda^{+}%
}{q+\lambda^{+}}(1-\mathbb{E}[e^{-(q+\lambda^{+})\tilde{\tau
}_{\varepsilon}%
}]) ) +o(1)}{1-\frac{\lambda^{+}}{q+\lambda^{+}}(1-\mathbb{E}%
[e^{-(q+\lambda^{+})\tilde{\tau}_{\varepsilon}}])+o(1)}.
\end{eqnarray}

We first consider $\tilde{X}$ has unbounded variation. From Lemma~\ref
{lem W},
we deduce that $\frac{W^{(q+\lambda^{+})\prime}(\varepsilon
)}{W^{(q+\lambda
^{+})}(\varepsilon)}\rightarrow\infty$ as $\varepsilon\downarrow
0$. By Theorem~\ref{thm asym}, this further implies that
%
\begin{equation}
1-\mathbb{E} \bigl[e^{-(q+\lambda^{+})\tilde{\tau}_{\varepsilon}} \bigr]=o(1). \label{o1}%
\end{equation}
One concludes from (\ref{joint}) and (\ref{o1}) that the denominator
on the
right-hand side of (\ref{qq}) approaches $1$ as $\varepsilon
\downarrow0$.
Moreover, by Theorem~\ref{thm asym},
\[
\lim_{\varepsilon\downarrow0}\frac{W^{(q+\lambda^{+})\prime
}(\varepsilon
)}{W^{(q+\lambda^{+})}(\varepsilon)} \bigl(1-\mathbb{E}
\bigl[e^{-q\tau
_{\varepsilon
}-sY_{\tau_{\varepsilon}}} \bigr] \bigr)=\lim_{\varepsilon\downarrow0}
\frac
{W^{(q+\lambda
^{+})\prime}(\varepsilon)}{W^{(q+\lambda^{+})}(\varepsilon)} \bigl( 1-\mathbb{E} \bigl[e^{-(q+\lambda^{+})\tilde{\tau}_{\varepsilon}-s\tilde
{Y}%
_{\tilde{\tau}_{\varepsilon}}} \bigr] \bigr) =s.
\]

When $\tilde{X}$ has bounded variation but the L\'{e}vy measure $\Pi
(-\infty,0)=\infty$, note that (\ref{o1}) still holds by Lemma~\ref
{lem W}.
Hence, it follows from (\ref{joint}) that the denominator on the right-hand
side of (\ref{qq}) also approaches $1$ as $\varepsilon\downarrow0$.
Furthermore, by Theorem~\ref{thm asym}, we obtain
\[
\lim_{\varepsilon\downarrow0}\frac{W^{(q+\lambda^{+})\prime
}(\varepsilon
)}{W^{(q+\lambda^{+})}(\varepsilon)} \bigl(1-\mathbb{E}
\bigl[e^{-q\tau
_{\varepsilon
}-sY_{\tau_{\varepsilon}}} \bigr] \bigr)=s+\frac{q+\lambda^{+}-\tilde{\psi
}(s)}{d}%
-
\frac{\lambda^{+}}{q+\lambda^{+}}\frac{q+\lambda^{+}}{d}=s+\frac
{q-\tilde{\psi}(s)}{d}.
\]

Finally, when $\tilde{X}$ has bounded variation and $\Pi(-\infty
,0)<\infty$,
by Lemma~\ref{lem W} and Theorem~\ref{thm asym},
%
\begin{equation}
\lim_{\varepsilon\downarrow0} \bigl( 1-\mathbb{E} \bigl[e^{-q\tilde{\tau
}_{\varepsilon}-s\tilde{Y}_{\tilde{\tau}_{\varepsilon}}} \bigr]
\bigr) =\frac{q+sd-\tilde{\psi}(s)}{q+\Pi(-\infty,0)}. \label{acp}%
\end{equation}
Then, by (\ref{qq}), (\ref{joint}) and Theorem~\ref{thm asym}, it is
straightforward to verify that
\[
\lim_{\varepsilon\downarrow0}\frac{W^{(q+\lambda^{+})\prime
}(\varepsilon
)}{W^{(q+\lambda^{+})}(\varepsilon)} \bigl(1-\mathbb{E}
\bigl[e^{-q\tau
_{\varepsilon
}-sY_{\tau_{\varepsilon}}} \bigr] \bigr)=s+\frac{q-\tilde{\psi}(s)}{d},
\]
which completes the proof.
\end{pf}

\section{Duration of drawdowns}
\label{sec:ddd}

In this section, we examine the duration of drawdowns via the LT of the
stopping time $\eta_{b}$ defined in (\ref{eta b}) for the L\'{e}vy
model with
two-sided jumps (\ref{Xph}).

To do so, we use the perturbation approach which has been developed by many
researchers in similar contexts (e.g., Dassios and Wu \cite{DW10}, Landriault
\textit{et al.} \cite{LRZ11}, Li and Zhou \cite{LZ13}, Loeffen \textit{et al.} \cite
{LCP13}, and
Zhang \cite{Z15}). To present the main idea, let $\varepsilon>0$ and define
the following sequence of stopping times:
\[
\tau_{\varepsilon}^{1}=\tau_{\varepsilon}, \qquad\vartheta
_{0}^{1}%
=\tau_{\varepsilon}^{1}+T_{M_{\tau_{\varepsilon}^{1}}}^{+}
\circ \theta _{\tau_{\varepsilon}^{1}},\ldots, \tau_{\varepsilon
}^{i}=
\vartheta _{0}^{i}+\tau_{\varepsilon}\circ
\theta_{\vartheta_{0}^{i}},\qquad %
\vartheta_{0}^{i}=
\tau_{\varepsilon}^{i}+T_{M_{\tau_{\varepsilon
}^{i}}}%
^{+}
\circ\theta_{\tau_{\varepsilon}^{i}},
\]
for $i\in%
\mathbb{N}
$ where we recall $\theta$ stands for the Markov shift operator. An
approximation of $\eta_{b}$ is given by
\[
\eta_{b}^{\varepsilon}=\inf \bigl\{ t\in(\tau_{\varepsilon
}^{i},
\vartheta _{0}^{i}]:t-\tau_{\varepsilon}^{i}
\geq b\mbox{ for some }i\in%
\mathbb{N} 
 \bigr\} ,
\]
for which only excursions of $Y$ with height over $\varepsilon$ are
considered. By construction, it is clear that $\eta_{b}^{\varepsilon
}$ is
monotonically decreasing as $\varepsilon\downarrow0$, and $\eta_{b}%
=\lim_{\varepsilon\downarrow0}\eta_{b}^{\varepsilon}$, $\mathbb{P}$-a.s.

For fixed $q>0$, we consider an independent exponential rv $\mathbf{e}_{q}$
with mean $1/q$. By the strong Markov property of $X$,
\begin{eqnarray*}
\mathbb{P} \bigl\{ \mathbf{e}_{q}>\eta_{b}^{\varepsilon}
\bigr\} & =&\mathbb{P} \bigl\{ \mathbf{e}_{q}>\eta_{b}^{\varepsilon},
\vartheta _{0}%
^{1}>\tau_{\varepsilon}^{1}+b
\bigr\} +\mathbb{P} \bigl\{ \mathbf {e}_{q}%
>
\eta_{b}^{\varepsilon},\vartheta_{0}^{1}<
\tau_{\varepsilon
}^{1}+b \bigr\}
\\
& =&\mathbb{P} \bigl\{ \mathbf{e}_{q}\wedge\vartheta_{0}^{1}>
\tau _{\varepsilon}^{1}+b \bigr\} +\mathbb{P} \bigl\{
\vartheta_{0}^{1}%
<\mathbf{e}_{q}
\wedge \bigl( \tau_{\varepsilon}^{1}+b \bigr) \bigr\} \mathbb{P} \bigl
\{ \mathbf{e}_{q}>\eta_{b}^{\varepsilon} \bigr\} ,
\end{eqnarray*}
which yields
%
\begin{equation}
\mathbb{P} \bigl\{ \mathbf{e}_{q}>\eta_{b}^{\varepsilon}
\bigr\} =\frac{\mathbb{P} \{ \mathbf{e}_{q}\wedge\vartheta
_{0}^{1}>\tau
_{\varepsilon}^{1}+b \} }{1-\mathbb{P} \{ \vartheta_{0}%
^{1}<\mathbf{e}_{q}\wedge ( \tau_{\varepsilon}^{1}+b )
 \}
}. \label{alp}%
\end{equation}
By conditioning on $Y_{\tau_{\varepsilon}^{1}}$ and then using the strong
Markov property of $X$ at time $\tau_{\varepsilon}^{1}$, we find
%
\begin{eqnarray}\label{alk1}%
\mathbb{P} \bigl\{ \vartheta_{0}^{1}<
\mathbf{e}_{q}\wedge \bigl( \tau_{\varepsilon}^{1}+b
\bigr) \bigr\} & =&\int_{[\varepsilon
,\infty
)}\mathbb{E} \bigl[
e^{-q\tau_{\varepsilon}}1_{ \{ Y_{\tau
_{\varepsilon}%
}\in\mathrm{d}y \} } \bigr] \mathbb{P} \bigl\{ T_{y}^{+}<
\mathbf{e}%
_{q}\wedge b \bigr\}
\nonumber
\\[-8pt]
\\[-8pt]
\nonumber
& =&\mathbb{E} \bigl[ e^{-q\tau_{\varepsilon}} \bigr] -\int_{[\varepsilon
,\infty)}
\mathbb{E} \bigl[ e^{-q\tau_{\varepsilon}}1_{ \{
Y_{\tau
_{\varepsilon}}\in\mathrm{d}y \} } \bigr] \mathbb{P} \bigl\{
T_{y}%
^{+}>\mathbf{e}_{q}\wedge b
\bigr\}
\end{eqnarray}
and
%
\begin{eqnarray} \label{alp1}%
\mathbb{P} \bigl\{ \mathbf{e}_{q}\wedge\vartheta_{0}^{1}>
\tau _{\varepsilon
}^{1}+b \bigr\} & =&\int_{[\varepsilon,\infty)}
\mathbb{E} \bigl[ e^{-q\tau_{\varepsilon}}1_{ \{ Y_{\tau_{\varepsilon}}\in
\mathrm{d}%
y \} } \bigr] \mathbb{P} \bigl\{
\mathbf{e}_{q}\wedge T_{y}%
^{+}>b
\bigr\}
\nonumber
\\[-8pt]
\\[-8pt]
\nonumber
& =&e^{-qb}\int_{[\varepsilon,\infty)}\mathbb{E} \bigl[
e^{-q\tau
_{\varepsilon}}1_{ \{ Y_{\tau_{\varepsilon}}\in\mathrm
{d}y \}
} \bigr] \mathbb{P} \bigl\{ T_{y}^{+}>b
\bigr\}.
\end{eqnarray}
Substituting (\ref{alk1}) and (\ref{alp1}) into (\ref{alp}), we obtain
%
\begin{equation}
\mathbb{E} \bigl[e^{-q\eta_{b}^{\varepsilon}} \bigr]=\mathbb{P} \bigl\{ \mathbf
{e}%
_{q}>\eta_{b}^{\varepsilon} \bigr\} =
\frac{e^{-qb}\int_{[\varepsilon,\infty
)}\mathbb{E} [ e^{-q\tau_{\varepsilon}}1_{ \{ Y_{\tau
_{\varepsilon}%
}\in\mathrm{d}y \} } ] \mathbb{P} \{
T_{y}^{+}>b \}
}{1-\mathbb{E} [ e^{-q\tau_{\varepsilon}} ] +\int_{[\varepsilon
,\infty)}\mathbb{E} [ e^{-q\tau_{\varepsilon}}1_{ \{
Y_{\tau
_{\varepsilon}}\in\mathrm{d}y \} } ] \mathbb{P} \{
T_{y}%
^{+}>\mathbf{e}_{q}\wedge b \} }. \label{mid3}%
\end{equation}
From the representation (\ref{mid3}), it seems relevant to define, for $x>0$
and $p\geq0$, a bounded auxiliary function
%
\begin{eqnarray} \label
{f eps}%
f_{\varepsilon}^{(p)}(t) &:=&\int_{[\varepsilon,\infty)}\mathbb {E}
\bigl[ e^{-q\tau_{\varepsilon}}1_{ \{ Y_{\tau_{\varepsilon}}\in
\mathrm{d}%
y \} } \bigr] \mathbb{P} \bigl\{
T_{y}^{+}>\mathbf {e}_{p}\wedge t \bigr\}
\nonumber
\\[-8pt]
\\[-8pt]
\nonumber
& =&\int_{[\varepsilon,\infty)}\mathbb{E} \bigl[ e^{-q\tau
_{\varepsilon}%
}1_{ \{ Y_{\tau_{\varepsilon}}\in\mathrm{d}y \} }
\bigr] \mathbb{P} \{ M_{\mathbf{e}_{p}\wedge t}\leq y \} ,
\end{eqnarray}
where the dependence of (\ref{f eps}) on $q$ is silently assumed.
Hence, we
rewrite (\ref{mid3}) as
%
\begin{equation}
\mathbb{E} \bigl[e^{-q\eta_{b}^{\varepsilon}} \bigr]=\frac
{e^{-qb}f_{\varepsilon}%
^{(0)}(b)}{1-\mathbb{E}[e^{-q\tau_{\varepsilon}}]+f_{\varepsilon}^{(q)}(b)}.
\label{mid4}%
\end{equation}

To obtain a well-defined asymptotics for $f_{\varepsilon}^{(p)}(b)$ as
$\varepsilon\downarrow0$, the key is to investigate the convergence
of the
measure $\mathbb{E} [ e^{-q\tau_{\varepsilon}}1_{ \{
Y_{\tau
_{\varepsilon}}\in\mathrm{d}y \} } ] $ as $\varepsilon
\downarrow
0$, which is closely related to the asymptotic results of Section~\ref
{sec:asym}. As we will see below, the convergence of the measure differs
according to whether the L\'{e}vy process has bounded or unbounded variation.

\subsection{Bounded variation case}
\label{sec:bv}

We first show that the distribution function of the running maximum of
$X$ is well-behaved.

\begin{proposition}
\label{prop bv}Let $X$ be a L\'{e}vy process of bounded variation with
a drift
$d>0$ in its characteristic exponent representation (\ref{bv cha}).
Then, for
any fixed $p\geq0$ and $t>0$, the function $\mathbb{P}\{M_{\mathbf{e}
_{p}\wedge t}\leq y\}/y$ is bounded for $y\in(0,\infty)$. Moreover,
if we
further assume that $\Pi(-\infty,0)=\infty$ and $\Pi$ has no atoms on
$(-\infty,0)$, the function $\mathbb{P}\{M_{\mathbf{e}_{p}\wedge
t}\leq y\}/y$
is also continuous for every $y\in(0,\infty)$.
\end{proposition}

\begin{pf}
For any fixed $p\geq0$ and $t>0$, we denote by
\[
F_{t}^{(p)}(y):=\mathbb{P}\{M_{\mathbf{e}_{p}\wedge t}\leq y\}/y.
\]
We first consider the case $p=0$. Using the upper bound in equation
(4.16) of
Chaumont and Ma{\l}ecki \cite{CM13} (which holds for a general L\'{e}vy process),
we know that
%
\begin{equation}
F_{t}^{(0)}(y)\leq\frac{e}{e-1}\kappa \biggl(
\frac{1}{t},0 \biggr) \frac
{h(y)}{y}, \label{F0 bd}%
\end{equation}
where we recall $h(\cdot)$ is the renewal function defined in (\ref{renewal}).
Since $X$ has bounded variation and $d>0$, we deduce that $X$ creeps upwards
by Theorem~7.11 of Kyprianou \cite{K06}. From Lemma~\ref{lem H} we
know that
$h(y)/y$ converges to a finite limit as $y\downarrow0$. Therefore, we conclude
from (\ref{F0 bd}) that $F_{t}^{(0)}(y)$ is bounded for $y\in
(0,\infty)$.

Next, we consider the case $p>0$. By Wiener--Hopf factorization, it is
well known that $\tilde{M}_{\mathbf{e}_{p}}$ follows an exponential
distribution with mean $1/\tilde{\Phi}(p)>0$. Moreover, since
$M_{t}\geq
\tilde{M}_{t}$ a.s. for any $t\geq0$, one obtains that
\begin{eqnarray*}
F_{t}^{(p)}(y) & =&\int_{(0,t)}pe^{-ps}
\frac{\mathbb{P}\{M_{s}\leq y\}
}%
{y}\,\mathrm{d}s+e^{-pt}\frac{\mathbb{P}\{M_{t}\leq y\}}{y}
\\
& \leq&\int_{(0,\infty)}pe^{-ps}\frac{\mathbb{P}\{\tilde
{M}_{s}\leq y\}}%
{y}
\,\mathrm{d}s+e^{-pt}F_{t}^{(0)}(y)
\\
& =&\frac{1-e^{-\tilde{\Phi}(p)y}}{y}+e^{-pt}F_{t}^{(0)}(y).
\end{eqnarray*}
By the boundedness of $F_{t}^{(0)}(\cdot)$, we deduce that
$F_{t}^{(p)}(y)$ is
also bounded for $y\in(0,\infty)$.

Finally, suppose that we also have $\Pi(-\infty,0)=\infty$ and $\Pi
$ has no
atoms on $(-\infty,0)$. For any fixed $t>0$, by Theorem~27.7 of Sato
\cite{Sato99}, we know that the law of $\tilde{X}_{t}$ is absolute continuous
with respect to the Lebesgue measure, so is the law of $X_{t}$ from the
property of convolutions. In addition, by Theorem~6.5 of Kyprianou
\cite{K06},
we know $X$ is regular for $(0,\infty)$ as $X$ has bounded variation and
$d>0$. Therefore, from Theorem~1 of Chaumont \cite{C13}, we conclude
the law
of $M_{t}$ is absolute continuous with respect to the Lebesgue measure.
As a
consequence, $\mathbb{P}\{M_{\mathbf{e}_{p}\wedge t}\leq y\}/y$ is continuous
for every $y\in(0,\infty)$.
\end{pf}

\begin{remark}
\label{rk AC}For the L\'{e}vy model (\ref{Xph}) with $\tilde{X}$ has bounded
variation and $\Pi(-\infty,0)=\infty$, it follows that $\mathbb{P}%
\{M_{\mathbf{e}_{p}\wedge t}\leq y\}/y$ is bounded and continuous for
$y\in(0,\infty)$ due to our assumptions that $|\tilde{X}|$ is not a
subordinator, $\tilde{X}$ is regular for $(0,\infty)$, and $\Pi$ has
no atom
on $(-\infty,0)$.
\end{remark}

We are now ready to present the main result of this subsection.

\begin{theorem}
\label{thm BV}Consider the L\'{e}vy model (\ref{Xph}). If $\tilde
{X}$ has
bounded variation and satisfies Assumption~\ref{assume1}, for any
$q>0$, we
have
\[
\mathbb{E} \bigl[e^{-q\eta_{b}} \bigr]=e^{-qb}\frac{\int_{(0,\infty)}\mathbb
{P}\{M_{b}\leq
y\}\Pi(-\mathrm{d}y)}{q+\int_{(0,\infty)}\mathbb{P}\{M_{\mathbf
{e}_{q}\wedge
b}\leq y\}\Pi(-\mathrm{d}y)}.
\]
\end{theorem}

\begin{pf}
We first consider the case $\Pi(-\infty,0)=\infty$. From (\ref{f
eps}) with
$p\geq0$, we have
%
\begin{eqnarray} \label{WF}%
&&\frac{W^{(q+\lambda^{+})\prime}(\varepsilon)}{W^{(q+\lambda^{+})}%
(\varepsilon)}f_{\varepsilon}^{(p)}(b)\nonumber\\
&&\quad  =\frac{W^{(q+\lambda
^{+})\prime
}(\varepsilon)}{W^{(q+\lambda^{+})}(\varepsilon)}\int
_{[\varepsilon
,\infty
)}\mathbb{E} \bigl[e^{-q\tau_{\varepsilon}}1_{\{Y_{\tau_{\varepsilon
}}\in
\mathrm{d}y\}}
\bigr] \mathbb{P}\{M_{\mathbf{e}_{p}\wedge b}\leq y\}
\nonumber
\\[-8pt]
\\[-8pt]
\nonumber
&&\quad =\int_{(0,\infty)}\frac{\mathbb{P}\{M_{\mathbf{e}_{p}\wedge
b}\leq
y\}}{1-e^{-y}}\cdot\frac{W^{(q+\lambda^{+})\prime}(\varepsilon)}%
{W^{(q+\lambda^{+})}(\varepsilon)}1_{\{y\geq\varepsilon\}}
\bigl(1-e^{-y}%
 \bigr)\mathbb{E} \bigl[e^{-q\tau_{\varepsilon}}1_{\{Y_{\tau_{\varepsilon}}\in
\mathrm{d}y\}}
\bigr]
\\
&&\quad =\int_{(0,\infty)}\frac{\mathbb{P}\{M_{\mathbf{e}_{p}\wedge
b}\leq
y\}}{1-e^{-y}}\mu_{\varepsilon}(
\mathrm{d}y),\nonumber
\end{eqnarray}
where $\mu_{\varepsilon}(\mathrm{d}y)$ is a finite measure on
$(0,\infty)$
defined as
%
\begin{equation}
\mu_{\varepsilon}(\mathrm{d}y)=\frac{W^{(q+\lambda^{+})\prime
}(\varepsilon
)}{W^{(q+\lambda^{+})}(\varepsilon)}1_{\{y\geq\varepsilon\}
}
\bigl(1-e^{-y}%
 \bigr)\mathbb{E} \bigl[e^{-q\tau_{\varepsilon}}1_{\{Y_{\tau_{\varepsilon}}\in
\mathrm{d}y\}}
\bigr]. \label{mumeasure}%
\end{equation}

By Theorem~\ref{thm asym2}, we have
\begin{eqnarray*}
\lim_{\varepsilon\downarrow0}\int_{(0,\infty)}e^{-sy}\mu
_{\varepsilon
}(\mathrm{d}y)&=&\lim_{\varepsilon\downarrow0}\frac{W^{(q+\lambda
^{+})\prime
}(\varepsilon)}{W^{(q+\lambda^{+})}(\varepsilon)}
\bigl(\mathbb {E} \bigl[e^{-q\tau
_{\varepsilon}-sY_{\tau_{\varepsilon}}} \bigr]-\mathbb{E} \bigl[e^{-q\tau
_{\varepsilon
}-(s+1)Y_{\tau_{\varepsilon}}}
\bigr] \bigr)\\
&=&1+\frac{\tilde{\psi}(s)-\tilde
{\psi}(s+1)}%
{d},
\end{eqnarray*}
for all $s\geq0$. On the other hand, we notice from (\ref{psi bv}) that
\begin{eqnarray*}
&&\int_{(0,\infty)}e^{-sy}\frac{1-e^{-y}}{d}\Pi(-
\mathrm{d}y)\\
&&\quad =\frac{1}%
{d}\int_{(-\infty,0)} \bigl(
e^{sy}-1 \bigr) \Pi(\mathrm {d}y)-\frac{1}{d}%
\int
_{(-\infty,0)} \bigl(e^{(s+1)y}-1 \bigr)\Pi(\mathrm{d}y)
\\
&&\quad =1+\frac{\tilde{\psi}(s)-\tilde{\psi}(s+1)}{d}.
\end{eqnarray*}
Hence, by Proposition~\ref{prop weak}, one concludes that, as
$\varepsilon
\downarrow0$, $\mu_{\varepsilon}(\mathrm{d}y)$ weakly converges to
the measure
$d^{-1}(1-e^{-y})\Pi(-\mathrm{d}y)$, which is a finite measure on
$(0,\infty)$
because $\tilde{X}$ has bounded variation.

From Proposition~\ref{prop bv} and Remark~\ref{rk AC}, we know the function
$\mathbb{P}\{M_{\mathbf{e}_{p}\wedge b}\leq y\}/(1-e^{-y})$ is
bounded and
continuous for $y\in(0,\infty)$. By the definition of weak
convergence, it
follows from (\ref{WF}) that
%
\begin{eqnarray}\label{lim bv}%
\lim_{\varepsilon\downarrow0}\frac{W^{(q+\lambda^{+})\prime
}(\varepsilon
)}{W^{(q+\lambda^{+})}(\varepsilon)}f_{\varepsilon}^{(p)}(b)
& =&\lim_{\varepsilon\downarrow0}\int_{(0,\infty)}\frac{\mathbb{P}\{
M_{\mathbf{e}%
_{p}\wedge b}\leq y\}}{1-e^{-y}}
\mu_{\varepsilon}(\mathrm {d}y)
\nonumber
\\
& =&\int_{(0,\infty)}\frac{\mathbb{P}\{M_{\mathbf{e}_{p}\wedge
b}\leq
y\}}{1-e^{-y}}\frac{1}{d}
\bigl(1-e^{-y} \bigr)\Pi(-\mathrm{d}y)
\\
& =&\frac{1}{d}\int_{(0,\infty)}\mathbb{P}\{M_{\mathbf{e}_{p}\wedge
b}
\leq y\}\Pi(-\mathrm{d}y). \nonumber
\end{eqnarray}
Therefore, by (\ref{mid4}), (\ref{lim bv}) and Theorem~\ref{thm
asym2}, we
have
\begin{eqnarray*}
\mathbb{E} \bigl[e^{-q\eta_{b}} \bigr] & =&\frac{e^{-qb}\lim_{\varepsilon
\downarrow
0}\frac{W^{(q+\lambda^{+})\prime}(\varepsilon)}{W^{(q+\lambda^{+}%
)}(\varepsilon)}f_{\varepsilon}^{(0)}(b)}{\lim_{\varepsilon
\downarrow0}%
\frac{W^{(q+\lambda^{+})\prime}(\varepsilon)}{W^{(q+\lambda^{+})}%
(\varepsilon)} ( 1-\mathbb{E}[e^{-q\tau_{\varepsilon}}] )
+\lim_{\varepsilon\downarrow0}\frac{W^{(q+\lambda^{+})\prime
}(\varepsilon
)}{W^{(q+\lambda^{+})}(\varepsilon)}f_{\varepsilon}^{(q)}(b)}
\\
& =&\frac{e^{-qb}\int_{(0,\infty)}\mathbb{P}\{M_{b}\leq y\}\Pi
(-\mathrm{d}%
y)}{q+\int_{(0,\infty)}\mathbb{P}\{M_{\mathbf{e}_{q}\wedge b}\leq
y\}\Pi(-\mathrm{d}y)}.
\end{eqnarray*}

Finally, we consider the case that $\Pi(-\infty,0)<\infty$. By (\ref{acp}),
for any $s\geq0$, we have
\[
\lim_{\varepsilon\downarrow0}\int_{(0,\infty)}e^{-sy}
\mathbb {E} \bigl[e^{-q\tau
_{\varepsilon}}1_{\{Y_{\tau_{\varepsilon}}\in\mathrm{d}y\}} \bigr]=\frac
{\tilde
{\psi}(s)-sd+\Pi(-\infty,0)}{q+\Pi(-\infty,0)}=\int
_{(0,\infty
)}e^{-sy}%
\frac{\Pi(-\mathrm{d}y)}{q+\Pi(-\infty,0)}.
\]
By Proposition~\ref{prop weak}, we see that the measure $\mathbb{E}%
[e^{-q\tau_{\varepsilon}}1_{\{Y_{\tau_{\varepsilon}}\in\mathrm
{d}y\}}]$ weakly
converges to the measure $\Pi(-\mathrm{d}y)/(q+\Pi(-\infty,0))$ as
$\varepsilon\downarrow0$. Since $\mathbb{P}\{M_{\mathbf
{e}_{p}\wedge b}\leq
y\}$ is bounded and upper semi-continuous in $y\in(0,\infty)$, it
follows from
Portemanteau theorem of weak convergence that
%
\begin{eqnarray}\label{f up}%
\limsup_{\varepsilon\downarrow0}f_{\varepsilon}^{(p)}(b) & =&\limsup
_{\varepsilon\downarrow0}\int_{(0,\infty)}\mathbb{P} \{
M_{\mathbf{e}%
_{p}\wedge b}\leq y \} \mathbb{E} \bigl[e^{-q\tau_{\varepsilon}}%
1_{\{Y_{\tau_{\varepsilon}}\in\mathrm{d}y\}}
\bigr]
\nonumber
\\[-8pt]
\\[-8pt]
\nonumber
& \leq&\frac{1}{q+\Pi(-\infty,0)}\int_{(0,\infty)}\mathbb{P}\{
M_{\mathbf{e}%
_{p}\wedge b}\leq y\}\Pi(-\mathrm{d}y).
\end{eqnarray}
On the other hand, since $\mathbb{P}\{M_{\mathbf{e}_{p}\wedge b}<y\}$
is lower
semi-continuous in $y\in(0,\infty)$. By Portemanteau theorem again,
we have
%
\begin{eqnarray} \label{f down}%
\liminf_{\varepsilon\downarrow0}f_{\varepsilon}^{(p)}(b) & \geq&
\liminf_{\varepsilon\downarrow0}\int_{(0,\infty)}\mathbb{P} \{
M_{\mathbf{e}_{p}\wedge b}<y \} \mathbb{E} \bigl[e^{-q\tau
_{\varepsilon}%
}1_{\{Y_{\tau_{\varepsilon}}\in\mathrm{d}y\}} \bigr]
\nonumber
\\
& \geq&\frac{1}{q+\Pi(-\infty,0)}\int_{(0,\infty)}\mathbb{P} \{
M_{\mathbf{e}_{p}\wedge b}<y \} \Pi(-\mathrm{d}y)
\\
& =&\frac{1}{q+\Pi(-\infty,0)}\int_{(0,\infty)}\mathbb{P}\{
M_{\mathbf{e}%
_{p}\wedge b}\leq y\}\Pi(-\mathrm{d}y),\nonumber
\end{eqnarray}
where the last equality holds because $\Pi$ has no atom on $(-\infty
,0)$ and
$\mathbb{P} \{ M_{\mathbf{e}_{p}\wedge b}<y \} =\mathbb{P}
\{M_{\mathbf{e}_{p}\wedge b}\leq y\}$ for almost all $y>0$. By letting
$\varepsilon\downarrow0$ in each term of (\ref{mid4}) and using
(\ref{f up}),
(\ref{f down}) and (\ref{acp}), this completes the proof of Theorem~\ref{thm BV}.
\end{pf}

\subsection{Unbounded variation case}
\label{sec:ubv}

We now consider the unbounded variation case for which the following
assumption on the density of $X_{t}$ is made.

\begin{assumption}
\label{assume2}If $X$ has unbounded variation, we assume that the
density of
$X_{t}$, namely $p_{t}^{X}(x)$, is bounded for all $t>0$.
\end{assumption}

\begin{remark}
\label{rk assume2} We point out that Assumption~\ref{assume2} is
identical to
assumption~(H1) in Chaumont and Ma{\l}ecki \cite{CM13}, which is
equivalent to
the assumption that the characteristic function $e^{-t\Psi(\cdot)}\in
L^{2}(\mathbb{R})$, for all $t>0$. It is also clear that, if $X$ is a
spectrally negative L\'{e}vy process with unbounded variation and $Y$
is an
arbitrary L\'{e}vy process independent of $X$, then the sum $X+Y$ satisfies
Assumption~\ref{assume2} as long as $X$ does. Hence, examples of Levy
processes satisfying Assumption~\ref{assume2} include processes with
$\sigma>0$, or $\sigma=0$ and with a spectrally negative $\alpha
$-stable jump
distribution with $\alpha\in(1,2)$.
\end{remark}

The following proposition shows that, for a L\'{e}vy process with unbounded
variation satisfying Assumption~\ref{assume2}, the density of the running
maximum at $0+$ is well-behaved.

\begin{proposition}
\label{prop M density}Let $X$ be a L\'{e}vy process with unbounded variation
that creeps upwards and satisfies Assumption~\ref{assume2}. Then the running
maximum $M_{t}$ has a continuous density $p_{t}^{M}(\cdot)$ for every $t>0$
and further,
\[
\lim_{x\downarrow0}p_{t}^{M}(x)=
\frac{\bar{\nu}_{L}(t)+\kappa
(0,0)}{d_{H}}>0,
\]
where $\bar{\nu}_{L}(\cdot)$ is the tail of the jump measure of the ascending
ladder time process (see (\ref{nuT})).
\end{proposition}

\begin{pf}
From Lemma~\ref{lem H}, we know that the renewal density $h^{\prime}$
can be
chosen to be a continuous function with well-defined limit $h^{\prime
}(0)=\frac{1}{d_{H}}>0$. Since $X$ has unbounded variation, assumption~(H2) of
Chaumont and Ma{\l}ecki \cite{CM13} also holds. By Proposition~2 and
Theorem~1 of
Chaumont and Ma{\l}ecki \cite{CM13}, we know that $M_{t}$ has a continuous
density $p_{t}^{M}(x)$ for every $t>0$, and also,
\[
\lim_{x\downarrow0}\frac{p_{t}^{M}(x)}{h^{\prime}(x)}=d_{H}\lim
_{x\downarrow
0}p_{t}^{M}(x)=n(\zeta>t),
\]
where $n(\zeta>t)$ is the excursion measure of excursions with length over
$t>0$. From (6.11) and (6.14) of Kyprianou \cite{K06} (see also
Section IV.4
of Bertoin \cite{B96}), we know that
\[
n(\zeta>t)=\bar{\nu}_{L}(t)+\kappa(0,0)>0,
\]
which completes the proof.
\end{pf}

\begin{corollary}
\label{cor ubv}Under the conditions of Proposition~\ref{prop M
density}, for
any fixed $p\geq0$ and $t>0$, the function $\mathbb{P}\{M_{\mathbf{e}
_{p}\wedge t}\leq y\}/y$ is bounded and continuous for $y\in{}[
0,\infty)$,
where its value at $y=0$ is defined as the right limit
\[
\lim_{y\downarrow0}\frac{\mathbb{P}\{M_{\mathbf{e}_{p}\wedge
t}\leq y\}}%
{y}=\frac{1}{d_{H}} \biggl( \int
_{(0,t)}pe^{-ps}\bar{\nu }_{L}(s)
\,\mathrm{d}%
s+e^{-pt}\bar{\nu}_{L}(t)+\kappa(0,0)
\biggr).
\]
\end{corollary}

\begin{pf}
From Proposition~\ref{prop M density}, it is only left to justify the
limit of
$\mathbb{P}\{M_{\mathbf{e}_{p}\wedge t}\leq y\}/y$ as $y\downarrow
0$. By
dominated convergence theorem and Proposition~\ref{prop M density}
again, we
have
\begin{eqnarray*}
\lim_{y\downarrow0}\frac{\mathbb{P}\{M_{\mathbf{e}_{p}\wedge
t}\leq y\}}{y} & =&\int_{0}^{t}pe^{-ps}
\lim_{y\downarrow0}\frac{\mathbb{P}\{
M_{s}\leq y\}}%
{y}\,\mathrm{d}s+e^{-pt}\lim
_{y\downarrow0}\frac{\mathbb{P}\{
M_{t}\leq y\}}{y}
\\
& =&\int_{0}^{t}pe^{-ps}\lim
_{y\downarrow0}p_{s}^{M}(y)\,\mathrm
{d}s+e^{-pt}%
\lim_{y\downarrow0}p_{t}^{M}(y)
\\
& =&\frac{1}{d_{H}} \biggl( \int_{(0,t)}pe^{-ps}
\bar{\nu }_{L}(s)\,\mathrm{d}%
s+e^{-pt}\bar{
\nu}_{L}(t)+\kappa(0,0) \biggr) ,
\end{eqnarray*}
which ends the proof.
\end{pf}

Now we are ready to present the main result of this subsection.

\begin{theorem}
\label{thm UNV}Consider the L\'{e}vy model (\ref{Xph}). If $\tilde
{X}$ has
unbounded variation and satisfies Assumptions \ref{assume1} and \ref{assume2},
for any $q>0$, we have
\[
\mathbb{E} \bigl[e^{-q\eta_{b}} \bigr]=e^{-qb}\frac{\bar{\nu}_{L}(b)+\kappa
(0,0)}%
{\int_{(0,b)}qe^{-qt}\bar{\nu}_{L}(t)\,\mathrm{d}t+e^{-qb}\bar{\nu}%
_{L}(b)+\kappa(0,0)}.
\]
\end{theorem}

\begin{pf}
It is clear that the L\'{e}vy model (\ref{Xph}) creeps upward as its
spectrally negative component $\tilde{X}$ does and its upward
jumps follow a compound Poisson structure. Moreover, since
$\tilde{X}$
satisfies Assumption~\ref{assume2}, by Remark~\ref{rk assume2}, we
see that
all the conditions of Proposition~\ref{prop M density} are satisfied.

For the finite measure $\mu_{\varepsilon}(\mathrm{d}y)$ defined in
(\ref{mumeasure}), it is straightforward to verify from Theorem~\ref{thm asym2} that, for any $s\geq0$,
\begin{eqnarray*}
\lim_{\varepsilon\downarrow0}\int_{\mathbb{R}_{+}}e^{-sy}\mu
_{\varepsilon
}(\mathrm{d}y)&=&\lim_{\varepsilon\downarrow0}\frac{W^{(q+\lambda
^{\prime
})\prime}(\varepsilon)}{W^{(q+\lambda^{+})}(\varepsilon)}
\bigl(\mathbb {E}%
 \bigl[e^{-q\tau_{\varepsilon}-sY_{\tau_{\varepsilon}}} \bigr]-\mathbb {E}
\bigl[e^{-q\tau
_{\varepsilon}-(s+1)Y_{\tau_{\varepsilon}}} \bigr] \bigr)\\
&=&1=\int_{\mathbb
{R}_{+}}%
e^{-sy}
\delta_{0}(\mathrm{d}y).
\end{eqnarray*}
It follows from Proposition~\ref{prop weak} that $\mu_{\varepsilon}%
(\mathrm{d}y)$ weakly converges to the Dirac measure $\delta
_{0}(\mathrm{d}y)$
as $\varepsilon\downarrow0$. Moreover, by Corollary~\ref{cor ubv},
we know
that the function $\mathbb{P}\{M_{\mathbf{e}_{p}\wedge t}\leq y\}/(1-e^{-y})$
is also bounded and continuous for $y\in{}[0,\infty)$, where its
value at
$y=0$ is defined by the limit as $y\downarrow0$. From (\ref{WF}) and Corollary~\ref{cor ubv}, we have
%
\begin{eqnarray}\label{stepb2}%
\lim_{\varepsilon\downarrow0}\frac{W^{(q+\lambda^{+})\prime
}(\varepsilon
)}{W^{(q+\lambda^{+})}(\varepsilon)}f_{\varepsilon}^{(p)}(b)
& =&\lim_{\varepsilon\downarrow0}\int_{\mathbb{R}_{+}}\frac{\mathbb
{P}%
\{M_{\mathbf{e}_{p}\wedge b}\leq y\}}{1-e^{-y}}
\mu_{\varepsilon}%
(\mathrm{d}y)
\nonumber
\\
& =&\lim_{y\downarrow0}\frac{\mathbb{P}\{M_{\mathbf{e}_{p}\wedge
b}\leq
y\}}{1-e^{-y}}
\\
& =&\frac{1}{d_{H}} \biggl( \int_{(0,b)}pe^{-pt}
\bar{\nu }_{L}(t)\,\mathrm{d}%
t+e^{-pb}\bar{
\nu}_{L}(b)+\kappa(0,0) \biggr). \nonumber
\end{eqnarray}
It follows from (\ref{mid4}), (\ref{stepb2}) and Theorem~\ref{thm asym2} that
\begin{eqnarray*}
\mathbb{E} \bigl[e^{-q\eta_{b}} \bigr] & =&e^{-qb}\frac{\lim_{\varepsilon
\downarrow
0}\frac{W^{(q+\lambda^{+})\prime}(\varepsilon)}{W^{(q+\lambda^{+}%
)}(\varepsilon)}f_{\varepsilon}^{(0)}(b)}{\lim_{\varepsilon
\downarrow0}%
\frac{W^{(q+\lambda^{+})\prime}(\varepsilon)}{W^{(q+\lambda^{+})}%
(\varepsilon)} ( 1-\mathbb{E}[e^{-q\tau_{\varepsilon
}}]+f_{\varepsilon
}^{(q)}(b) ) }
\\
& =&e^{-qb}\frac{\bar{\nu}_{L}(b)+\kappa(0,0)}{\int_{(0,b)}qe^{-qt}\bar{\nu
}_{L}(t)\,\mathrm{d}t+e^{-qb}\bar{\nu}_{L}(b)+\kappa(0,0)},
\end{eqnarray*}
which ends the proof.
\end{pf}

In general, the function $\bar{\nu}_{L}$ and $\kappa(0,0)$ are only
implicitly
known via (\ref{nuT}) and Wiener--Hopf factorization. When $X$ has no positive
jumps, we can express $\mathbb{E}[e^{-q\eta_{b}}]$ explicitly in
terms of
$p_{t}^{X}$.

\begin{corollary}
\label{cor Kendall}Let $X$ be a spectrally negative L\'{e}vy process with
unbounded variation and satisfies Assumptions \ref{assume1} and \ref{assume2}.
For any $q>0$, we have
\[
\mathbb{E} \bigl[e^{-q\eta_{b}} \bigr]=e^{-qb}\frac{\int_{(b,\infty)}\frac
{1}{s}p_{s}%
^{X}(0)\,\mathrm{d}s}{\int_{(0,b)}qe^{-qt}\int_{(t,\infty)}\frac
{1}{s}p_{s}%
^{X}(0)\,\mathrm{d}s\,\mathrm{d}t+e^{-qb}\int_{(b,\infty)}\frac
{1}{s}p_{s}%
^{X}(0)\,\mathrm{d}s}.
\]
\end{corollary}

\begin{pf}
By Kendall's identity, for any fixed $t,y>0$, we have
\[
\frac{1}{y}\mathbb{P}\{M_{t}\leq y\}=\frac{1}{y}\int
_{(t,\infty)}%
\mathbb{P} \bigl\{T_{y}^{+}
\in\,\mathrm{d}s \bigr\}=\int_{(t,\infty)}\frac
{1}{s}p_{s}%
^{X}(y)
\,\mathrm{d}s.
\]
It follows from Fourier inversion that, for any $y\in\mathbb{R}$ and $s>0$,
\[
0\leq\frac{1}{s}p_{s}^{X}(y)\leq\frac{1}{2\pi s}
\int_{\mathbb{R}}%
\bigl|e^{-s\Psi(u)}\bigr|\,\mathrm{d}u.
\]
From the proof of Proposition~5 of Chaumont and Ma{\l}ecki \cite{CM13},
we know
that for any fixed $t>0$,
\[
\frac{1}{2\pi}\int_{(t,\infty)}\frac{1}{s}\int
_{\mathbb
{R}}\bigl|e^{-s\Psi
(u)}\bigr|\,\mathrm{d}u\,\mathrm{d}s<\infty.
\]
By the dominated convergence theorem and Corollary~\ref{cor ubv}, we have
\[
\frac{\bar{\nu}_{L}(t)+\kappa(0,0)}{d_{H}}=\lim_{y\downarrow
0}\frac
{\mathbb{P}\{M_{t}\leq y\}}{y}=\int
_{(t,\infty)}\frac{1}{s}p_{s}%
^{X}(0)
\,\mathrm{d}s,
\]
which completes the proof by using Theorem~\ref{thm UNV}.
\end{pf}

\section{Examples}
\label{sec:ex}

\begin{example}
Consider a Brownian motion, i.e. $X_{t}=\mu t+\sigma W_{t}$ with
$\sigma>0$.
For any fixed $t>0$, we have
\[
p_{t}^{X}(x)=\frac{1}{\sigma\sqrt{2\pi t}}\exp \biggl\{ -
\frac
{(x-\mu t)^{2}%
}{2\sigma^{2}t} \biggr\}.
\]
By Remarks \ref{rk assume1}, \ref{rk assume2} and Corollary~\ref{cor
Kendall},
we have
\[
\mathbb{E} \bigl[e^{-q\eta_{b}} \bigr]=\frac{e^{-qb}g(b)}{\int_{(0,b)}qe^{-qt}%
g(t)\,\mathrm{d}t+e^{-qb}g(b)},
\]
where $g(t):=\int_{(t,\infty)}\frac{1}{s}p_{s}^{X}(0)\,\mathrm
{d}s=\frac
{2}{\sigma\sqrt{2\pi t}}e^{-{\mu^{2}t}/{(2\sigma^{2})}}-\frac
{2\mu}{\sigma
}N(-\frac{\mu\sqrt{t}}{\sigma})$ and $N(\cdot)$ is the cumulative
distribution
function of a standard normal rv.
\end{example}

\begin{example}
Consider a spectrally negative $\alpha$-stable process with Laplace exponent
$\psi(s)=s^{\alpha}$ with $\alpha\in(1,2)$. For fixed $t>0$, it is
well known
(e.g., pages 87--88 of Sato \cite{Sato99}) that
\[
p_{t}^{X}(x)=\frac{1}{\pi}t^{-{1}/{\alpha}}\sum
_{n=1}^{\bolds
{\infty}%
}(-1)^{n-1}
\frac{\Gamma(1+{n}/{\alpha})}{n!}\sin \biggl( \frac
{n\pi}%
{\alpha} \biggr) \bigl(t^{-{1}/{\alpha}}x
\bigr)^{n-1},
\]
where $\Gamma(\cdot)$ is the Gamma function. It follows that
\[
\int_{(t,\infty)}\frac{1}{s}p_{s}^{X}(0)
\,\mathrm{d}s=\frac{\alpha
}{\pi}%
\Gamma \biggl( \frac{1}{\alpha}
\biggr) \sin \biggl( \frac{\pi
}{\alpha} \biggr) t^{-{1}/{\alpha}}.
\]
By Remarks \ref{rk assume1}, \ref{rk assume2} and Corollary~\ref{cor
Kendall},
we have
\[
\mathbb{E} \bigl[e^{-q\eta_{b}} \bigr]=\frac{1}{e^{qb}b^{{1}/{\alpha}}\int_{(0,b)}qe^{-qt}t^{-{1}/{\alpha}}\,\mathrm{d}t+1}.
\]
\end{example}

\begin{example}
Consider a spectrally negative Gamma process with Laplace exponent
\[
\psi(s)=sd+\int_{(-\infty,0)} \bigl(e^{sx}-1 \bigr)
\beta|x|^{-1}e^{\alpha x}%
\,\mathrm{d}x=sd-\beta\log(1+s/
\alpha),\qquad s\in\mathbb{H}^{+},
\]
where $\alpha,\beta>0$ are constants. From Remark~\ref{rk char}, we define
\[
\varphi(s):=\frac{s\psi^{\prime}(s)-\psi(s)}{\psi(s)^{2}}=\frac
{\beta
\log(1+s/\alpha)-{\beta s}/{(s+\alpha)}}{ ( sd-\beta\log
(1+s/\alpha
) ) ^{2}}.
\]
One can easily verify that, for any fixed $s_{0}>\Phi(0)$, we have
$\varphi(s_{0}+{i}\cdot)\in L^{1}(\mathbb{R})$ which implies
Assumption~\ref{assume1} holds. Using Kendall's identity and
\[
p_{t}^{X}(x)=\frac{\alpha^{\beta s}}{\Gamma(\beta t)}(sd-y)^{\beta
s-1}e^{-\alpha(sd-y)}1_{\{y<sd\}},\qquad
x\in%
\mathbb{R} 
, t>0,
\]
we have
\[
\mathbb{P} \bigl\{ T_{y}^{+}>t \bigr\} =y\int
_{(t,\infty)}\frac
{1}{s}%
\frac{\alpha^{\beta s}}{\Gamma(\beta s)}1_{\{y<sd\}}(sd-y)^{\beta
s-1}e^{-\alpha(sd-y)}
\,\mathrm{d}s,\qquad  y>0\mbox{ and }t>0.
\]
Hence, by Fubini's theorem followed by some calculations, we can show that
\begin{eqnarray*}
&&\int_{(0,\infty)}\mathbb{P} \{ M_{t}\leq y \} \Pi (-
\mathrm{d}y) \\
&&\quad =\int_{(0,\infty)}\mathbb{P} \bigl\{
T_{y}^{+}>t \bigr\} \Pi (-\mathrm{d}y)
\\
& &\quad=\int_{(0,\infty)}y\int_{(t,\infty)}
\frac{1}{s}\frac{\alpha
^{\beta s}%
}{\Gamma(\beta s)}1_{\{y<sd\}}(sd-y)^{\beta s-1}e^{-\alpha(sd-y)}%
\,\mathrm{d}s\,\beta\frac{e^{-\alpha y}}{y}\,\mathrm{d}y
\\
&&\quad =(d\alpha)^{\beta s}\int_{(t,\infty)}\frac{1}{\Gamma(\beta
s)}s^{\beta
s-2}e^{-\alpha sd}
\,\mathrm{d}s.
\end{eqnarray*}
Using Theorem~\ref{thm BV}, one concludes
\[
\mathbb{E} \bigl[e^{-q\eta_{b}} \bigr]=\frac{e^{-qb}(d\alpha)^{\beta s}\int_{(b,\infty
)}\frac{s^{\beta s-2}e^{-\alpha sd}}{\Gamma(\beta s)}\,\mathrm{d}s}%
{q+(d\alpha)^{\beta s}\int_{(0,b)}qe^{-qt}\,\mathrm{d}t\int_{(t,\infty)}%
\frac{s^{\beta s-2}e^{-\alpha sd}}{\Gamma(\beta s)}\,\mathrm
{d}s+e^{-qb}%
(d\alpha)^{\beta s}\int_{(b,\infty)}\frac{s^{\beta s-2}e^{-\alpha
sd}}%
{\Gamma(\beta s)}\,\mathrm{d}s}.
\]
\end{example}

\begin{example}
Consider Kou's jump-diffusion model given by
\[
X_{t}=\mu t+\sigma W_{t}+\sum
_{i=1}^{N_{t}^{+}}J_{i}^{+}-\sum
_{j=1}^{N_{t}%
^{-}}J_{j}^{-},
\]
where $\mu\in%
\mathbb{R}
$, $\sigma>0$, $N^{\pm}$ are two independent Poisson processes with arrival
rates $\lambda^{\pm}>0$ and $J^{\pm}$ are a sequence of i.i.d. exponentially
distributed random variables with mean $1/\eta^{\pm}>0$. Its Laplace exponent
is given by
\[
\psi(s)=\frac{\sigma^{2}}{2}s^{2}+\mu s+\lambda^{-} \biggl(
\frac
{\eta^{-}%
}{\eta^{-}+s}-1 \biggr) +\lambda^{+} \biggl( \frac{\eta^{+}}{\eta
^{+}%
-s}-1
\biggr) ,\qquad  s\in \bigl(-\eta^{-},\eta^{+} \bigr).
\]
According to Corollary~1 of Asmussen \textit{et al.} \cite{AAP04} and
Section~6.5.4 of
Kyprianou \cite{K06}, it is known that the Laplace exponent of the ascending
ladder height is given by
\[
\kappa(\alpha,\beta)=\frac{(\beta+\rho_{1,\alpha})(\beta+\rho
_{2,\alpha}%
)}{(\beta+\eta^{+})}, \qquad \alpha,\beta\geq0,
\]
where $\rho_{1,\alpha}$ and $\rho_{2,\alpha}$ (with $\rho
_{1,\alpha}<\eta
^{+}<\rho_{2,\alpha}$) are the two distinct nonnegative solutions of
\mbox{$\psi(s)=\alpha$.} By Remarks \ref{rk assume1}, \ref{rk assume2} and Theorem~\ref{thm UNV}, one obtains
\[
\mathbb{E} \bigl[e^{-q\eta_{b}} \bigr]=e^{-qb}\frac{\bar{\nu}_{L}(b)+\frac
{\rho_{1,0}%
\rho_{2,0}}{\eta^{+}}}{\int_{(0,b)}qe^{-qt}\bar{\nu}_{L}(t)\,\mathrm
{d}%
t+e^{-qb}\bar{\nu}_{L}(b)+\frac{\rho_{1,0}\rho_{2,0}}{\eta^{+}}}.
\]
\end{example}

\begin{appendix}
\section*{Appendix}
\label{sec:app}

The following result is from Theorem~5.22 of Kallenberg \cite{K02}.

\begin{theorem}[(Extended continuity theorem)]\label{thm.A} Let $\mu_{1},\mu
_{2},\ldots$ be
probability measures on $\mathbb{R}^{d}$ with characteristic functions
$\hat{\mu}_{n}(t)\rightarrow\varphi(t)$ pointwisely for every $t\in
\mathbb{R}^{d}$, where the limit $\varphi$ is continuous at 0. Then
$\mu_{n}$
converges weakly to $\mu$ for some probability measure $\mu$ in
$\mathbb{R}%
^{d}$ with $\hat{\mu}=\varphi$. A corresponding statement holds for the
Laplace transforms of measures on $\mathbb{R}_{+}^{d}$.
\end{theorem}

\begin{proposition}
\label{prop weak}Let $\{\mu_{n}\}_{n\in%
\mathbb{N}
}$ be finite measures on $[0,\infty)$ with Laplace transforms
\[
\hat{\mu}_{n}(s)=\int_{\mathbb{R}_{+}}e^{-sy}
\mu_{n}(\,\mathrm{d}y),
\]
for $n\in%
\mathbb{N}
$ and $s\geq0$. Suppose that $\lim_{n\rightarrow\infty}\hat{\mu}%
_{n}(s)=\varphi(s)$ for all $s\geq0$, where $\varphi(\cdot)$ is a
positive and
continuous function on $[0,\infty)$. Then $\mu_{n}$ weakly converges
to $\mu$
as $n\rightarrow\infty$, for some finite measure $\mu$ on $[0,\infty
)$, and
$\hat{\mu}=\varphi$.
\end{proposition}

\begin{pf}
Since $\lim_{n\rightarrow\infty}\hat{\mu}_{n}(0)=\varphi(0)>0$,
we can
consider a sequence of probability measures $\nu_{n}(\mathrm
{d}y):=\mu
_{n}(\mathrm{d}y)\hat{\mu}_{n}(0)^{-1}$. By our assumptions, it is
easy to
see
\[
\lim_{n\rightarrow\infty}\hat{\nu}_{n}(s)=\varphi(s)
\varphi(0)^{-1},
\]
which is a continuous function at $0$. By Theorem~\ref{thm.A}, one concludes
that $\{\nu_{n}\}_{n\in%
\mathbb{N}
}$ weakly converges to some probability measure $v$ on $[0,\infty)$ with
$\hat{v}(\cdot)=\varphi(\cdot)\varphi(0)^{-1}$. Therefore, by letting
$\mu(\cdot):=v(\cdot)\varphi(0)$, we can see that $\mu_{n}$ weakly
converges
to $\mu$ as $n\rightarrow\infty$ and $\hat{\mu}=\varphi$.
\end{pf}
\end{appendix}


%





\printhistory
\end{document}